\documentclass[pss]{wiley2sp} % provides pss two-column style
\usepackage{amsmath}

\usepackage{amssymb}
\usepackage{amsmath}
\usepackage{siunitx}

\newcommand{\MGe}{\mathcal{G}^{(e)}}

\newcommand{\MWe}{W^{(e)}}

\newcommand{\Vdc}{V_{\rm dc}}
\newcommand{\Vac}{V_{\rm ac}}

\newcommand{\Fmu}{|F_0\rangle}
\newcommand{\tbar}{\overline{t}}
\newcommand{\Sbar}{\overline{S}}
\newcommand{\Qbar}{\overline{Q}}
\newcommand{\qbar}{\overline{q}}

\begin{document}

% Title of the article
\title{Two-particle interferometry in quantum Hall edge channels}

% Abbreviated title for the page headers
\titlerunning{Two-particle interferometry in quantum Hall edge channels}

% Authors
\author{%
  A. Marguerite\textsuperscript{\textsf{\bfseries 1}},
  E. Bocquillon\textsuperscript{\Ast,\textsf{\bfseries 1}},
  J.-M. Berroir\textsuperscript{\textsf{\bfseries 1}},
  B. Pla\c cais\textsuperscript{\textsf{\bfseries 1}},
  A. Cavanna\textsuperscript{\textsf{\bfseries 2}},
  Y. Jin\textsuperscript{\textsf{\bfseries 2}},
  P. Degiovanni\textsuperscript{\textsf{\bfseries 3}},
  G. F\`eve\textsuperscript{\Ast,\textsf{\bfseries 1}}
  }

% Abbreviated list of authors for the page headers
\authorrunning{A. Marguerite et al.}

%E-mail-address of corresponding author
\mail{e-mail
  \textsf{bocquillon@lpa.ens.fr, feve@lpa.ens.fr}, Phone:
  +33-1 44 32 25 79}

% author's affiliations/addresses
\institute{ \textsuperscript{1}\,
Laboratoire Pierre Aigrain, Ecole Normale Sup\'erieure-PSL Research University, CNRS, Universit\'e Pierre et Marie Curie-Sorbonne
Universit\'es, Universit\'e Paris Diderot-Sorbonne Paris Cit\'e, 24 rue Lhomond, 75231 Paris Cedex 05, France\\
 \textsuperscript{2} Centre de Nanosciences et Nanotechnologies - Campus de Marcoussis
CNRS, Universit\'e Paris-Sud, Universit\'e Paris-Saclay, Route de Nozay, 91460 Marcoussis, France\\
 \textsuperscript{3}\, Univ Lyon, Ens de Lyon, Universit\'e Claude Bernard Lyon 1, CNRS, Laboratoire de Physique, F-69342 Lyon, France
 }

%\received{XXXX, revised XXXX, accepted XXXX} % do not change, will be filled in by the publisher
%\published{XXXX} % do not change, will be filled in by the publisher

% Please select about four verbal keywords for your manuscript.
\keywords{Electron quantum optics, single electron sources, noise and correlations}

\abstract{%
% This is a macro for the typesetting of two-column text in an
% abstract. It will typeset the two arguments in \abstcol{}{} as the
% left and right column inside the abstract box. At the
% columnbreak there will be always a columnbreak (\par), so both
% columns start with a new paragraph. No automatic column height
% balancing is done.
%
% If used with a \titlefigure it will silently output both
% parameters as consecutive paragraphs.
%
% The macro is defined exclusively inside the argument of \abstract{};
% if used outside it will raise an error.
%
% Usage: \abstcol{<left column>}{<right column>}
\abstcol{%
  Since pioneering works of Hanbury-Brown and Twiss, intensity-intensity correlations have been widely used in astronomical systems, for example to detect binary stars. They reveal statistics effects and two-particle interference, and offer a decoherence-free probe of the coherence properties of light sources. In the quantum Hall edge channels, the concept of quantum optics can transposed to electrons, and an analogous two-particle interferometry can be developed, in order to characterize single-electron states. We review in this article the recent experimental and theoretical progress on this topic.  }}

\titlefigure{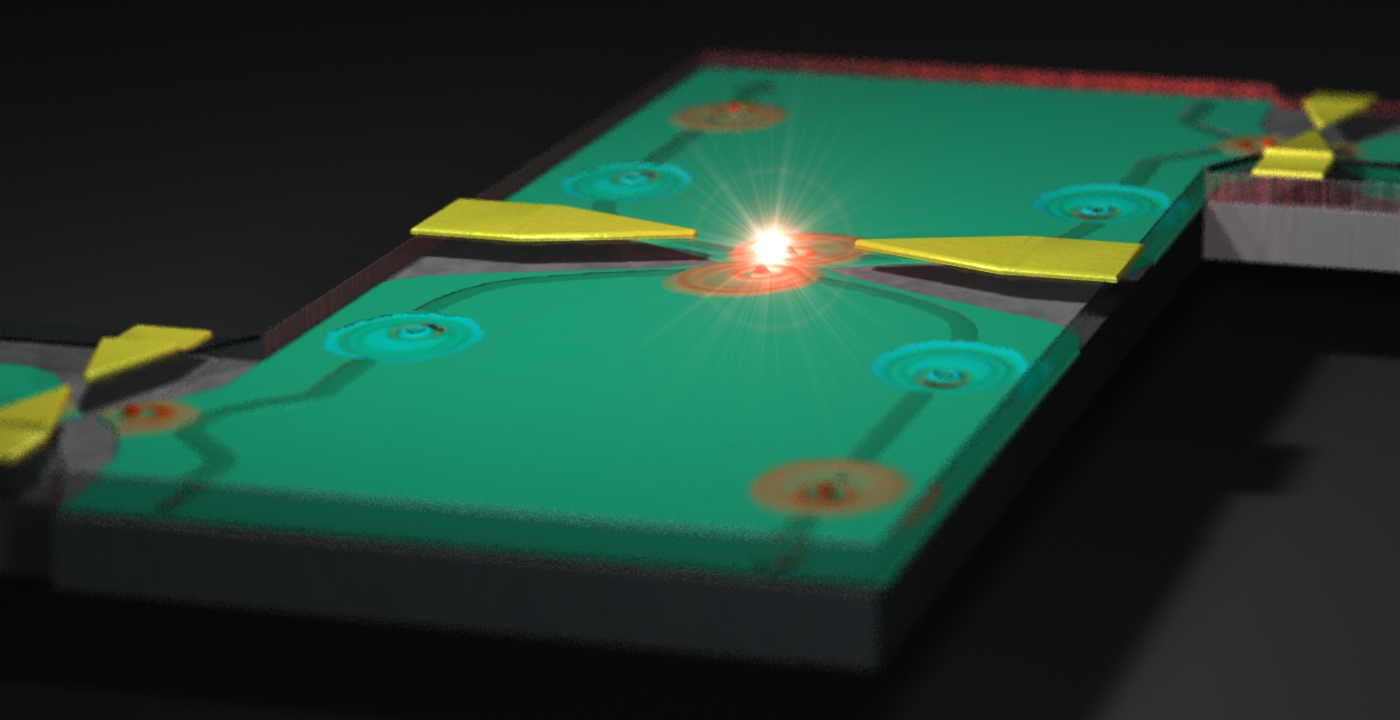}
\titlefigurecaption{}

\maketitle

\section{Introduction}
%Mesoscopic physics deals with the description of electronic transport in systems where electronic phase coherence or quantum statistics can not be neglected but rather play a prominent role. Such effects have in the first place been revealed in dc transport, in which the system is placed in a non-equilibrium stationary state. When interactions are not too strong, this regime can be entirely described by the Landauer-Buttiker scattering formalism

The manipulation of single to few electrons in a quantum conductor, directly relevant to the processing and transfer of information encoded in the electrical current, calls for a time-dependent description of electronic transport. Contrary to the dc regime, the non-stationary case requires the use of tools able to capture both the energetic and temporal aspects of electronic transport, for example using a time-dependent scattering formalism \cite{Moskalets2002,Moskalets2008}. When the elementary scale of single charge is reached, this formalism relates the measurement of electrical current to the wavefunctions of the particles propagating in the conductor.

In chiral edge channels of the quantum Hall effect, this wavefunction approach to electron transport (stationary or not) highlights the strong analogies with photons propagating in optical fibers. For example, (stationary) optics-like setups have been theoretically and experimentally investigated since the early 00s \cite{Henny1999,Oliver1999,Ji2003,Samuelsson2004,Splettstoesser2009}, in which electronic waves are produced in the quantum Hall edge channels and interfere. They yield information on how electrons propagate and interact with one another and with neighboring quasiparticles in the Fermi sea \cite{Neder2006,Roulleau2007,Levkivskyi2008}. In the same system, the recent development of on-demand single electron sources has allowed for time-controlled experiments at the single electron scale with increasing accuracy. In this context, it is possible to envision electron quantum optics experiments in which one controls the production, transfer and characterization of elementary quantum states in a conductor.

Most sources rely on the triggered release of single charges confined in quantum dots, such as the mesoscopic capacitor \cite{Buttiker1993,Gabelli2006a,Feve2007a,Gabelli2012}, or electron pumps or turnstiles \cite{Fricke2011,Mirovsky2013,Fletcher2013,Ubbelohde2014,Waldie2015}. Without confinement, the application of voltage pulses $V(t)$ directly on an ohmic contact can also generate single particles if they are designed with a Lorentzian shape and such that $e \int dt\,V(t)=h$ \cite{Levitov1996,Ivanov1997}, as experimentally verified \cite{Dubois2013b}. Thanks to accurate time control (in the \si{\pico\second} range), short single electron wavepackets (typically a few tens of \si{\pico\second}) can be precisely controlled to give rise to electron quantum optics experiments \cite{Bocquillon2014}, in which interference and correlations at the single particle scale can be studied to infer information on single particle states and on their evolution along propagation in the conductor, revealing for example single electron decoherence due to interaction with the environment \cite{Wahl2014,Ferraro2014,Slobodeniuk2016}.

Of particular interest, one can observe two-particle interference of independently emitted electrons. Though already detected with non-triggered dc excitations \cite{Liu1998,Neder2007}, such experiments take their full meaning when the two input arms of a beamsplitter are fed with synchronized clock-controlled single-electron wavepackets. They reveal effects such as indistinguishability, coherence (and decoherence) of quantum electronic states. In this article, we wish to review some of the recent progress in the use of two-particle interferometry in quantum Hall edge channels.

\begin{figure*}[htb]%
\includegraphics*[width=\textwidth]{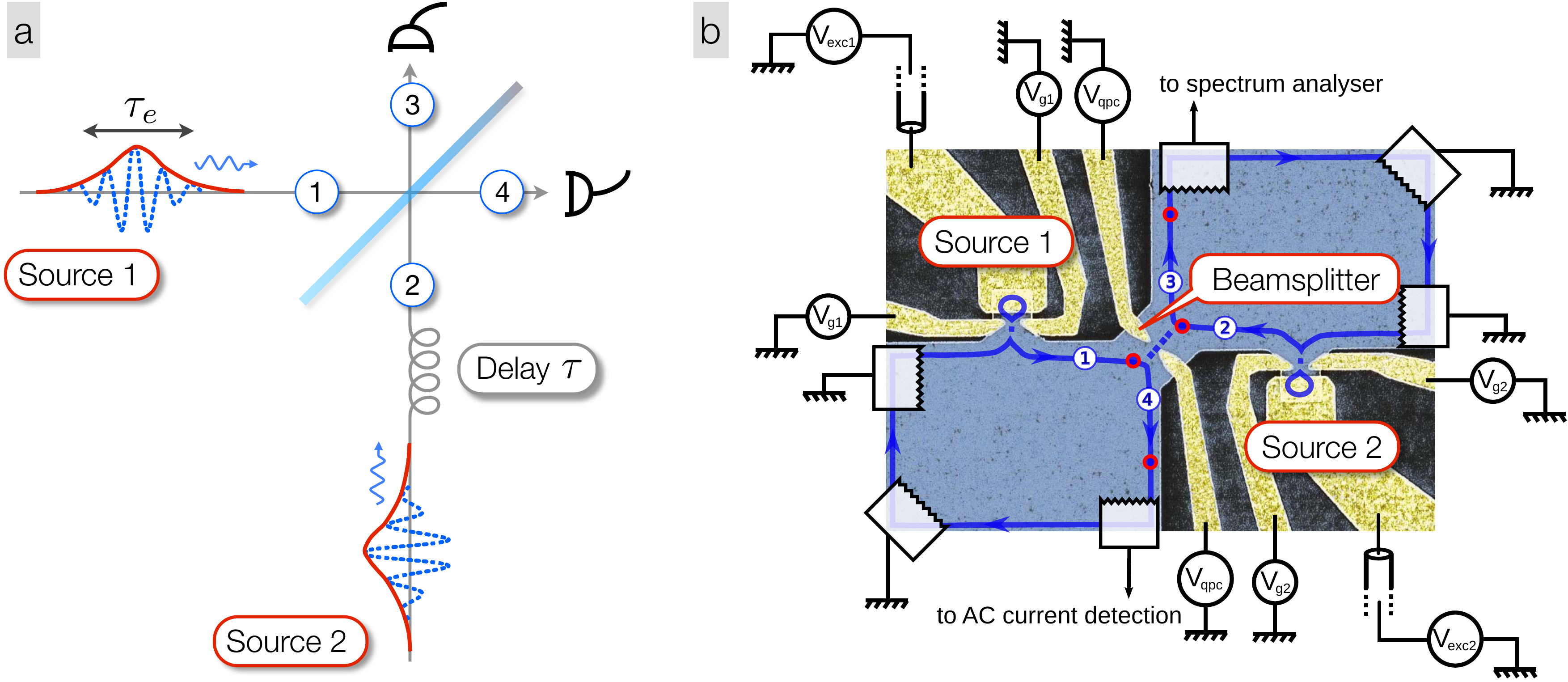}
\caption{a) Sketch of the Hong-Ou-Mandel geometry - Two beams of particles (source 1 and 2) impinge on the two input arms (1 and 2) of a beamsplitter. Cross-correlations or auto-correlations of output ports 3 and 4 reveal two-particle interference between both sources. A tunable delay $\tau$ can be additionally used to modulate the intensity of two-particle interference. b) Experimental realization in quantum Hall edge channels - two single electron sources (here based on the mesoscopic capacitor \cite{Feve2007a}) create single electron wavepackets that interfere on a quantum point contact, used as a beamsplitter. Current (auto-)correlations are measured in output 3 to implement two-particle interferometry protocols.}
\label{Fig:HOMSketch}
\end{figure*}

The review is organized as follows. In section \ref{Section:GeneralPrinciples}, we present the main experimental and theoretical concepts behind electron quantum optics, with a focus on two-particle interferometry. Section \ref{Section:HOM} concerns the two-particle interference between two identical but independent single electron sources. Finally, we describe in section \ref{Section:Tomo} a protocol to characterize a single electron source by means of two-particle interferometry.

\section{General principles of two-particle interferometry}
\label{Section:GeneralPrinciples}

\subsection{Two-particle interference : Hong-Ou-Mandel effect}

Both wave and particle aspects play an important role in the propagation of electrons. For example, interference fringes in the output current of a Mach-Zehnder interferometer \cite{Ji2003,Roulleau2007,Litvin2007} strikingly illustrates the wave nature of electrons. In contrast, current correlations in a shot noise experiment \cite{Henny1999,Oliver1999} reflect more prominently the corpuscular nature of the charge carriers. Yet, experiments involving the exchange statistics of indistinguishable \cite{Liu1998,Neder2007,Splettstoesser2009,Rossello2015} particles cannot be explained within merely wave or corpuscular descriptions, but require a full quantum treatment. As they ultimately deal with coherence and indistinguishability of electron quantum states, they are of particular interest to quantum information protocols in propagating electronic states in conductors \cite{Bertoni2000}.

\begin{figure}[htb]%
\includegraphics*[width=0.5\textwidth]{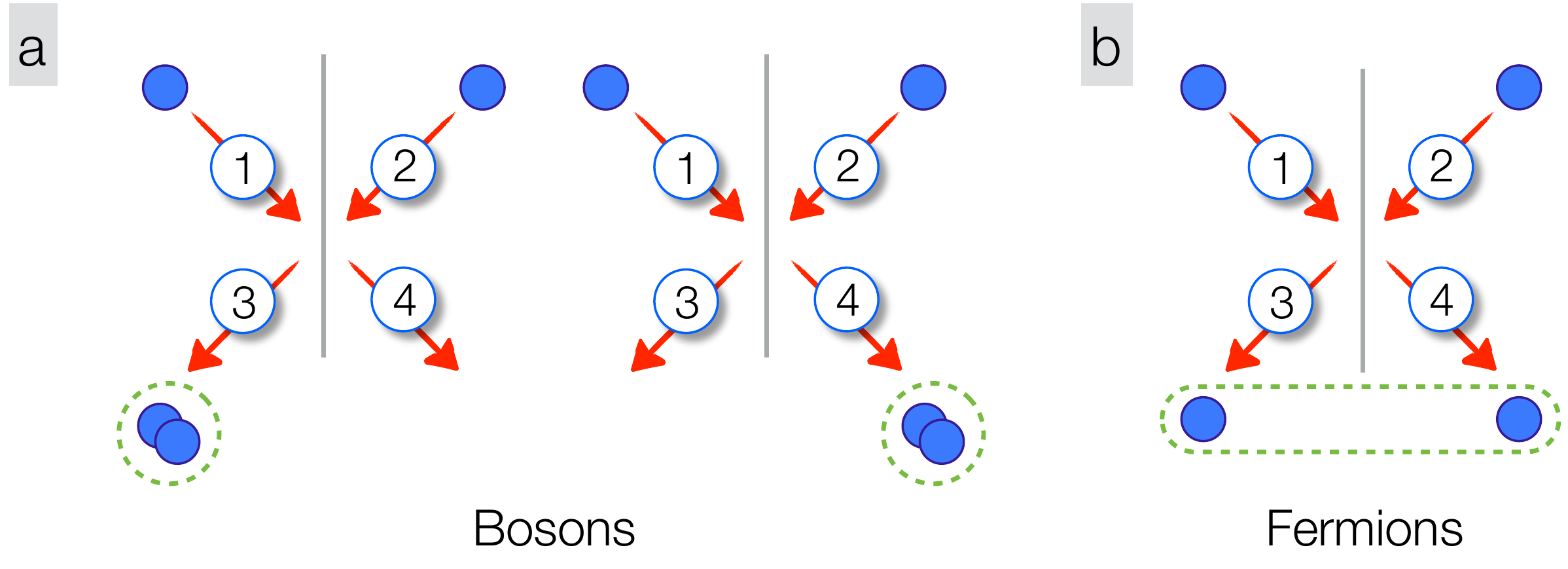}
\caption{Two-particle interference - When two indistinguishable particles reach the two inputs of a beamsplitter, quantum statistics enforces two-particle interference. a) Bosons (such as photons) bunch and always exit in the same output, with two possible outcomes. b) Fermions (such as electrons) antibunch, and always exit in different outputs, with one possible result.}
\label{Fig:HBTEffect}
\end{figure}

The Hong-Ou-Mandel (HOM) interferometer \cite{Hong1987}, sketched in Fig.\ref{Fig:HOMSketch}, is an example of experiment where exchange statistics effects are prominent. It comprises a beamsplitter, onto which two beams of particles collide and interfere (inputs 1 and 2). Measurements of the current correlations (auto- or cross-correlations) in the output ports (3 and 4) yield information on the exchange statistics of the two incoming particle. When two particles reach simultaneously a beamsplitter, four outcomes are a priori possible. In the seminal experiment realized in 1987 by Hong {\it et al.} \cite{Hong1987}, two photons were colliding on the splitter. In this case, Bose-Einstein statistics imposes that two indistinguishable bosons impinging simultaneously on a beamsplitter are not randomly partitioned, but rather "bunch" and always exit in the same output port, as depicted in Fig.\ref{Fig:HBTEffect}a. The cross-correlations between ports 3 and 4 then cancel out. A tunable delay $\tau$ between the input arms enables to progressively suppress the two-particle interference: when the impinging particles have no temporal overlap and thus do not interfere, one recovers the sum of the shot noises of the two inputs. Two-particle interferences thus appear as a cancellation of the cross-correlation around $\tau=0$, as a unique signature.
This first experiment, which used a source of twin photons, demonstrated the indistinguishability of the photons forming the pair. More recently, a couple of experiments have reproduced the same results with independent photon sources \cite{Beugnon2006,Flagg2010}.

In contrast, Fermi-Dirac statistics imposes that indistinguishable electrons would "antibunch" and exit in opposite outputs (Fig.\ref{Fig:HBTEffect}b). In this case, one expects the cross-correlation to reach a maximum in amplitude around $\tau=0$, and the auto-correlations to cancel out. It has recently been possible to implement such experiments in quantum Hall edge states \cite{Bocquillon2013a}. A scanning electron microscope picture of the device is presented in Fig.\ref{Fig:HOMSketch}b, and mimicks the HOM geometry. The HOM two-particle interference provides a probe of the degree of indistinguishability of two electron sources. Consequently a two-particle interferometer is a powerful tool to compare and characterize electronic states. Before moving to the applications of two-particle interferometry, we first introduce the theoretical framework to describe generic electronic states and how their coherence properties can be probed within the HOM geometry.

\subsection{Coherence and Wigner functions}

A new theoretical framework can be constructed relying on the analogies between photons and electrons in ballistic conductors. It conveniently describes the propagation of wavepackets containing a small number of electrons in a one-dimensional ballistic conductor, such as a quantum Hall edge channel. These theoretical tools rely on the formal analogy between the electric field operator $\hat E^{+}(x,t)$ (that annihilates photons at time $t$ and position $x$), and the electron field operator $\hat \Psi(x,t)$ (that annihilates electron at time $t$ and position $x$). In analogy with Glauber's theory of optical coherence \cite{Glauber1963}, the coherence of electron wavepackets can be investigated by defining coherence functions of first \cite{Grenier2011NJP,Haack2011,Haack2013} and second order \cite{Moskalets2014,Thibierge2016}, as well as a Wigner distribution function \cite{Ferraro2013}. We here briefly introduce the main tools, with a focus on the Wigner representation of single-particle coherence, and emphasize their application to two-electron interferometry in the next section.

In analogy with quantum optics, the coherence function (of degree one) is defined for electrons as\footnote{To simplify this equation, we drop the $x$-dependence, assuming a propagation at constant velocity, and $x=x'=0$.}
\begin{eqnarray}
\MGe(t,t')&=&\langle\hat\Psi^\dagger(t')\hat\Psi(t)\rangle
\end{eqnarray}
Beyond the analogies with optics, several specificities of the fermionic statistics have to be treated. At equilibrium, electrons form a Fermi sea $\Fmu$ (reference of chemical potential), whose coherence function $\MGe_{F_0}$ is non-zero in contrast with the photon vacuum. One can then decompose as:
\begin{eqnarray}
\MGe(t,t')&=&\MGe_{F_0}(t-t')+\Delta\MGe(t,t')\\
\MGe_{F_0}(t-t')&=&\int d\omega\, e^{i\omega (t-t')} f_0(\omega),
\end{eqnarray}
where $f_0(\omega)$ is the Fermi-Dirac electronic distribution function. It is important to notice that, since the Fermi sea is a stationary state, the coherence function of the Fermi sea only depends on the difference $t-t'$. On the opposite, dynamical states such as propagating wavepackets depend separately on $t$ and $t'$.

The coherence function in the time domain encodes all the relevant information on single particle transport, but it is often hard to exploit it and obtain direct physical insights on the propagating quantum state. It is often more convenient to use a mixed time-frequency representation, which encodes both temporal and energy aspects of the electronic state, by performing a Fourier transform with respect to $\tau=t-t'$. It defines the analog of the Wigner function for electrons \cite{Ferraro2013} similar to that of a particle \cite{Wigner1932} or an electric field \cite{Smithey1993,Bertet2002}.
\begin{eqnarray}
\MWe(\tbar,\omega)&=&\int d\tau\, \MGe(\tbar+\frac{\tau}{2}, \tbar-\frac{\tau}{2}) e^{i\omega\tau}
\end{eqnarray}
Given hermiticity properties of $\MGe$, $\MWe$ is a real function of $\tbar$ and $\omega$. As previously, we isolate the contribution of the Fermi sea $\Fmu$:
\begin{eqnarray}
\Delta\MWe(\tbar,\omega)&=&\MWe(\tbar,\omega)-\MWe_{F_0}(\omega)\label{Eq:DefDeltaW}\\
\MWe_{F_0}(\omega)&=&f_0(\omega)\label{Eq:DefWmu}
\end{eqnarray}
For a stationary state, the $\tbar$ dependence vanishes and the state is fully characterized by the electronic distribution function $f(\omega)$. This is no longer true in the time dependent case, where $\Delta\MWe(\tbar,\omega)$ depends explicitly on both $\tbar$ and $\omega$.
In full generality, the integration over $\tbar$ (resp. $\omega$) gives the probability distribution of the energy $\omega$ (resp. time $\tbar$). This naturally defines the excess particle distribution $\Delta f(\omega)$ created atop the Fermi sea as :
\begin{eqnarray}
\Delta f(\omega)&=&\frac{1}{T}\int_0^T d\tbar\, \Delta\MWe(\tbar,\omega)\label{Eq:DefDeltaF},
\end{eqnarray}
(where $T$ can either be the measurement time, or the period if the source is periodic) and the electrical current generated by the source:
\begin{eqnarray}
I(\tbar) & =& -e  \int \frac{d\omega}{2\pi }\Delta\MWe(\tbar,\omega)\label{Eq:DefI}
\end{eqnarray}
For classical states, $\MWe$ can be interpreted as a time-dependent electronic distribution function. This requires that, for all $\tbar$ and $\omega$, $0\leq\MWe(\tbar,\omega)\leq1$. Negative or above unity values are the hallmark of non-classical states.

\begin{figure}[t]%
\includegraphics*[width=\linewidth]{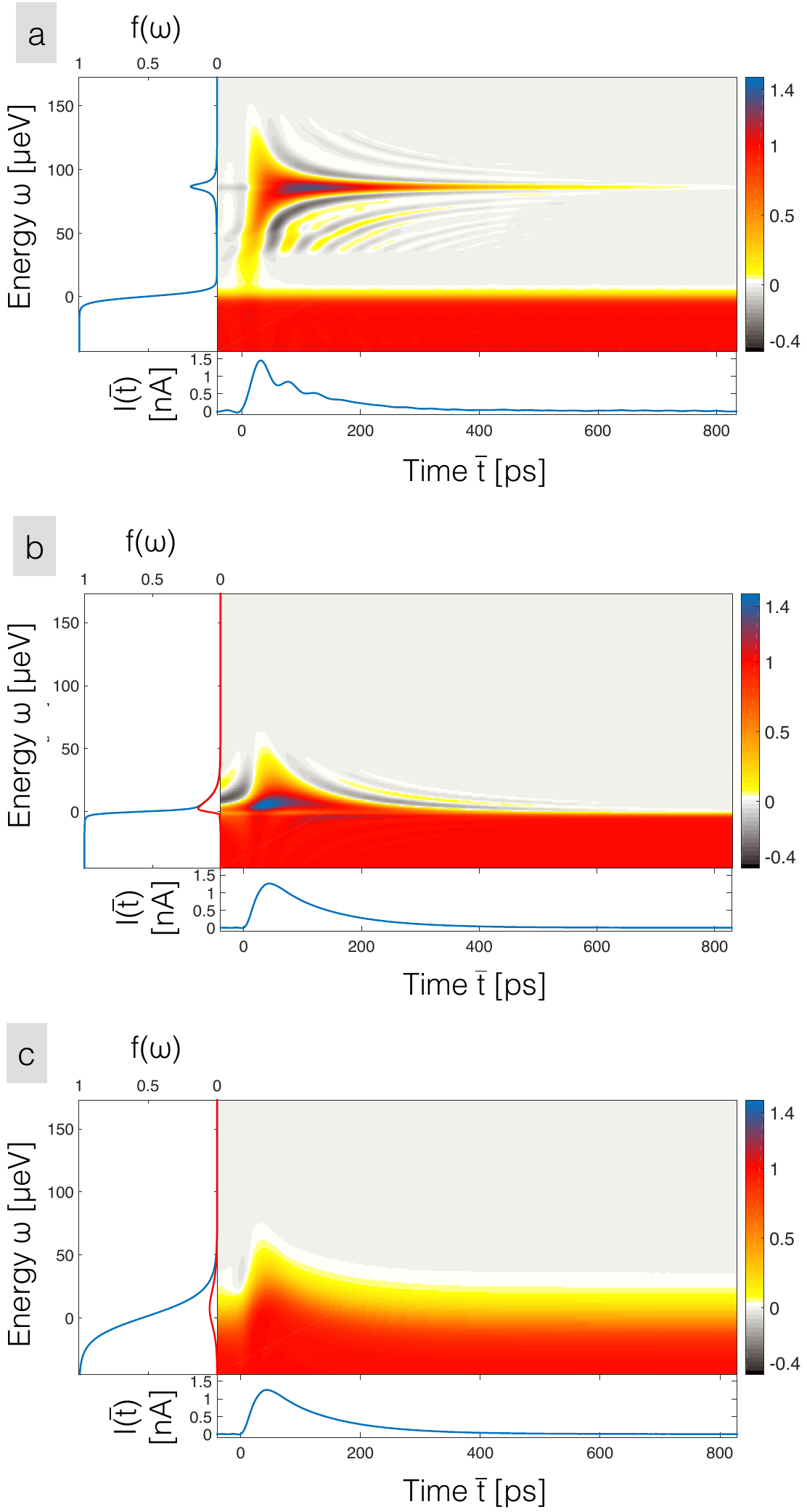}
\caption{Wigner functions corresponding to different exponential wavepackets - $\MWe$ is shown as a colorplot, in the central panel. In the lower panel, the current $I(t)$ obtained as the average of $\MWe$ over $\omega$ (Eq.\ref{Eq:DefI}) is shown as a blue line. On the left panel, the particle distribution $ f(\omega)$ obtained by averaging $\MWe$ over $\tbar$ (Eq.\ref{Eq:DefDeltaF}) is shown as the blue line. The difference $|\Delta f(\omega)|$ is shown as the red line. a) Energy resolved wavepacket generated above the Fermi sea, as predicted for the mesoscopic capacitor \cite{Ferraro2013}. Parameters: $\tau_0=\SI{110}{\pico\second}, T_{el}=\SI{25}{\milli\kelvin},\hbar\omega_0=\SI{85}{\micro\electronvolt}$. b) Wavepacket produced by driving an ohmic contact with an exponential drive, in the quantum limit $h/\tau_0\gg k_BT_{el}$ Parameters: $\tau_0=\SI{110}{\pico\second}, T_{el}=\SI{10}{\milli\kelvin}$. c) Wavepacket produced by driving an ohmic contact with an exponential drive, in the classical limit $h/\tau_0\ll k_BT_{el}$. Parameters: $\tau_0=\SI{110}{\pico\second}, T_{el}=\SI{100}{\milli\kelvin}$.}
\label{Fig:Wigner}
\end{figure}

These theoretical tools are very powerful to encompass any possible electronic state in a unique framework. It is particularly simple and insightful in the specific case of a single-electron wavepacket, such as the ones emitted by the lorentzian pulse source or by the mesoscopic capacitor. The emitted quantum states can be described as a single electron created in a wavepacket $\varphi^{(e)}(t)$, and we find the following expressions:
\begin{eqnarray}
\hat\Psi^\dagger[\varphi_e]\Fmu&=&\int dt\, \varphi_e(t)\hat\Psi^\dagger(t)\Fmu\\
\Delta\MGe(t,t')&=&\varphi_e(t)\varphi_e^*(t')\\
\Delta\MWe(\tbar,\omega)&=&\int d\tau\, \varphi_e(\tbar+\frac{\tau}{2})\varphi_e^*(\tbar-\frac{\tau}{2}) e^{i\omega\tau}
\end{eqnarray}
To illustrate these formulas, we present in Fig.\ref{Fig:Wigner} color plots of simulated Wigner functions obtained in the case of exponential wavepackets generated in different conditions\footnote{Other cases (lorentzian pulses, ac driven Fermi sea) are detailed in ref.\cite{Ferraro2013}.} Though the generated current $I(\tbar)$ (shown in the lower panels) is in all three cases exponentially decaying, the associated Wigner functions $\MWe(\tbar,\omega)$ and energy distributions $f(\omega)$ are very different (resp. central and left panels). This emphasizes the fact that different states (either classical and non-classical) can lead to the same current, but that the knowledge of their Wigner function enables to distinguish them. As a first example, the mesoscopic capacitor (Fig.\ref{Fig:Wigner}a) yields electrons flying above the Fermi sea, around an energy $\hbar\omega_0=\SI{85}{\micro\electronvolt}$, with an exponential decay time of $\tau_0=\SI{110}{\pico\second}$. The energy resolution around $\omega_0$ in enhanced for larger values of $\tbar$, as a manifestation of Heisenberg uncertainty principle.The associated quantum states is non-classical, as regions with $\MWe(\tbar,\omega)<0$ or $\MWe(\tbar,\omega)>1$ are observed. In contrast, a contact driven with an exponentially decaying voltage (Fig.\ref{Fig:Wigner}b and c) generates excitations close to the Fermi sea. If the typical energy scale $h/\tau_0$ is large against the electron temperature $T_{el}$ (i.e. $h/\tau_0\gg k_BT_{el}$, Fig.\ref{Fig:Wigner}b), the state is non-classical. If $h/\tau_0\ll k_BT_{el}$ (Fig.\ref{Fig:Wigner}c), the classical regime is reached, and $0<\MWe(\tbar,\omega)<1$ for all $\tbar$ and $\omega$. The state is then equivalent to a Fermi sea with varying chemical potential, with the Wigner function $\MWe(\tbar,\omega)=f_0\big(\omega+eV(t)/\hbar\big)$.

The coherence and Wigner functions are convenient representations of wavepackets containing one to a few electrons or holes. As their photonic counterparts, these quantities naturally appear in observables such as currents and current correlations. In the following section, we introduce this formalism to describe the Hong-Ou-Mandel geometry, and show how current and current correlations are related to the coherence and Wigner functions in this particular geometry.

\subsection{Principles of two-particle interferometry in the Hong-Ou-Mandel geometry}

The beamsplitter is described by energy-independent reflexion and transmission coefficients $\mathcal{R}$ and $\mathcal{T}$ (with $\mathcal{R}+\mathcal{T}=1$), connecting the input to the output ports. The current operator $\hat I_\alpha$ in output port $\alpha=3,4$ is given by $\hat I_\alpha(t)=e\,\hat\Psi^\dagger_\alpha(t)\hat\Psi_\alpha(t)$.
The current correlators $S_{\alpha\beta}(t,t')=\langle\delta \hat I_\alpha (t) \delta \hat I_\beta (t')$ can then be written as a function of the input currents and correlators:
\begin{eqnarray}
S_{33}&=&\mathcal{R}^2 S_{11}+\mathcal{T}^2 S_{22}+\mathcal{RT}\,Q\\
S_{44}&=&\mathcal{T}^2 S_{11}+R^2 S_{22}+\mathcal{RT}\,Q\\
S_{34}&=&\mathcal{R}\mathcal{T}\left(S_{11}+S_{22}-Q\right)
\end{eqnarray}
In this expression, $S_{11}(t,t')$ and $S_{22}(t,t')$ are simply the input noise in channels 1 and 2 transmitted through the beamsplitter. They can be measured in the absence of partitioning \cite{Mahe2010,Albert2010,Jonckheere2012a,Parmentier2012}. They encode the charge statistics of the sources, but are of no interest in the context of two particle interferometry. The last term $Q(t,t')$ encodes the partitioning terms, and contains the two-particle interference discussed in Section \ref{Section:GeneralPrinciples}.

In electronic transport, one in fact more easily accesses the time-averaged low-frequency noise $\Sbar_{\alpha\beta}$ defined as:
\begin{eqnarray}
\Sbar_{\alpha\beta}&=&2 \int d\tau\, \overline{S_{\alpha\beta} (\tbar+\tau/2,\tbar-\tau/2)}^{\tbar}
\end{eqnarray}
where $\overline{\cdots}^{\tbar}$ denotes an average over time $\tbar$.

The corresponding quantity $\Qbar$ is then easily recast in terms of the Wigner functions $\MWe_i$ at input $i$ of the beam-splitter :
\begin{eqnarray}
\Qbar&=&2 e^2 \int \frac{d\omega}{2\pi}\, \big[\overline{\MWe_1}^{\tbar} + \overline{\MWe_2}^{\tbar} -2\overline{\MWe_1 \MWe_2}^{\tbar}\big]\label{Eq:QbarHOM}
\end{eqnarray}
This equation embodies the main idea of two-particle interferometry, and the main message of this review article. The measurement of low-frequency noise yields the overlap of the Wigner function. It offers a way to characterize the first order coherence of the source, by directly comparing the two quantum states in the two input ports. One notes that the same information is in principle directly available at the output of a conventional one-particle interferometer such as a Mach-Zehnder interferometer \cite{Haack2011,Haack2013}. Though the device is more complex, the measurements of average currents at the output of a one-particle interferometer are far simpler and more accurate than measurements of correlations in a two-particle one. Yet, we argue that two-particle interferometers realize a "punctual" characterization device, while single-particle interferometers are by nature of finite length. Consequently, HOM interferometry is immune to decoherence effects in the measurement device, and directly probes the coherence punctually at the beamsplitter. In contrast, the response of Mach-Zehnder interferometers can be massively altered by interactions within the arms of the interferometer  \cite{Levkivskyi2008,Neder2007}.

Using Eqs.(\ref{Eq:DefDeltaW},\ref{Eq:DefWmu}), $\Qbar$ can be expanded as a sum of four terms, reading:
\begin{eqnarray}
\Qbar_{\rm eq}&=&4e^2 \int \frac{d\omega}{2\pi}\, f_{0}(\omega)\left(1-f_{0}(\omega)\right)\\
\Qbar_{\rm HBT,1}&=&2e^2\int \frac{d\omega}{2\pi}\,\overline{\Delta\MWe_{1}(\tbar,\omega)}^{\tbar}\left(1-2f_{0}(\omega)\right)\\
\Qbar_{\rm HBT,2}&=&2e^2\int \frac{d\omega}{2\pi}\, \overline{\Delta\MWe_{2}(\tbar,\omega)}^{\tbar}\left(1-2f_{0}(\omega)\right)\\
\Qbar_{\rm HOM}&=&-4e^2\int \frac{d\omega}{2\pi}\, \overline{\Delta\MWe_{1}(\tbar,\omega)\Delta\MWe_{2}(\tbar,\omega)}^{\tbar}
\end{eqnarray}
The physical meaning of each quantity is then very clear. $\Qbar_{eq}$ is the equilibrium contribution of both Fermi seas in ports 1 and 2. More interesting, $\Qbar_{{\rm HBT},i}$ represents the partition noise of source $i$ only (while the other source is switched off). In contrast with photons, the Fermi sea has in particular a non-trivial contribution $1-2f_0$, which represents two-particle interferences occurring on the beam-splitter between excitations emitted by source $i$ and the thermal excitations on the other input port. It can significantly modify the partition noise \cite{Bocquillon2012}, as will be recalled in Section \ref{Section:Tomo:HBT}. At zero temperature, this effect is suppressed and one recovers the random partitioning of classical particles on the splitter. Finally, at the core of this manuscript is the so called Hong-Ou-Mandel contribution $\Qbar_{\rm HOM}$, that records the two-particle interference between the two excitations generated in source 1 and 2 and measures the overlap between the excess Wigner functions. Two different cases will be envisioned in the following sections. In Section \ref{Section:HOM}, sources 1 and 2 are designed to be identical. The measurement of low-frequency noise then provides a measurement of their degree of indistinguishability. In a second case (Section \ref{Section:Tomo}), an unknown source in port 1 is compared to various reference sources in port 2 (biased Fermi sea, or sinusoidal density waves), in order to reconstruct the whole Wigner function $\Delta\MWe_1$ via a tomography protocol.

\section{Two-particle interference of identical sources}
\label{Section:HOM}
\subsection{Coherence and indistinguishability of electronic wavepackets}

A very natural experiment consists in placing two independent but identical single electron sources in the two input arms. It is convenient to define a normalized quantity $\Delta \qbar$ as:
\begin{eqnarray}
\Delta \qbar&=&\frac{\Qbar_{\rm HBT,1}+\Qbar_{\rm HBT,2}+\Qbar_{\rm HOM}}{\Qbar_{\rm HBT,1}+\Qbar_{\rm HBT,2}}\\
&=&1+\frac{\Qbar_{\rm HOM}}{\Qbar_{\rm HBT,1}+\Qbar_{\rm HBT,2}}
\end{eqnarray}
In this configuration, and for synchronized excitations, one expects to measure the perfect overlap of two identical states, so that $|\Qbar_{\rm HOM}|$ reaches its maximum value (namely $2\Qbar_{\rm HBT}$), and $\Delta\qbar=0$.
As the arrival time of the wavepackets is progressively shifted by a delay $\tau\neq0$, their overlap decays to 0, $\Qbar_{\rm HOM}=0$ and the full partition noise is recovered, with $\Delta \qbar=1$. This dip in the current auto-correlations is then analogous to the so-called HOM dip observed in light cross-correlation \cite{Hong1987}.
In the simple case of a wavepacket $\varphi_i$ in each arm $i=1,2$, one can show that $\Delta \qbar$ takes the simple form $\Delta \qbar=1-|\langle\varphi_1|\varphi_2\rangle|^2$, and $\Delta \qbar$ thus varies between 0 and 1.

At $\tau=0$, a perfect overlap can however only be obtained if the two incoming wavepackets are perfectly coherent and undistinguishable \cite{Olkhovskaya2008}. In this geometry, the contrast of two-particle interferences at $\tau=0$ is thus a direct indicator of the degree of indistinguishability of the excitations generated by the two sources.
Besides, the decay of the two-particle interference signal when the delay $\tau$ is increased provides additional information on the temporal shape of the wavepacket. An analogy can be drawn with photons, where the length of the photon pulse or radiative lifetime is reflected in the shape of the HOM dip \cite{Hong1987}.

\begin{figure}[htb]%
\includegraphics*[width=0.45\textwidth]{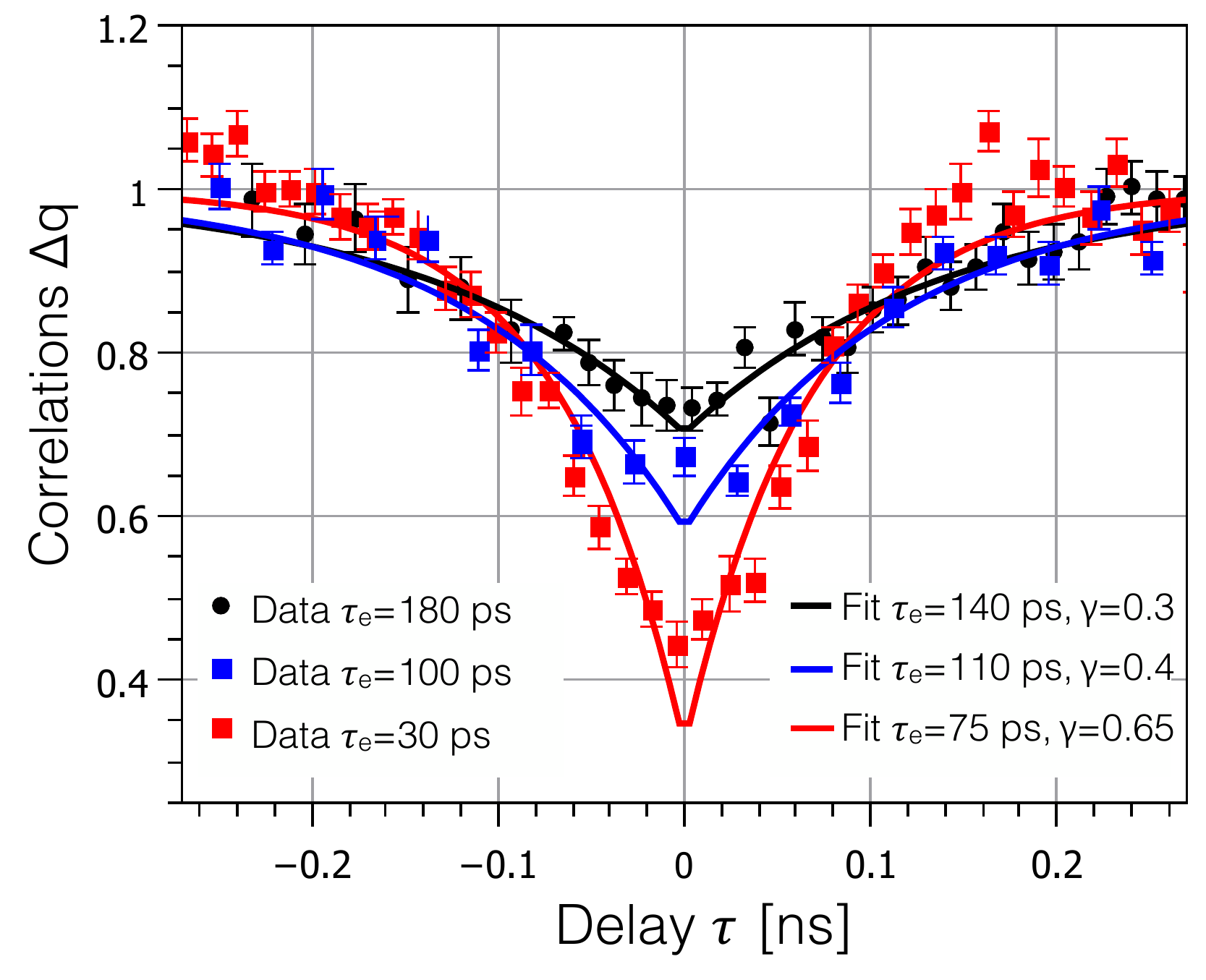}
\caption{HOM dips - Correlations $\Delta q$ as a function of time delay $\tau$, for three different values of the temporal width $\tau_e$. The measured data is presented as symbols with error bar. Fits with an exponential decay $\Delta q(\tau)=1-\gamma e^{-|\tau|/\tau_e}$ are shown as solid lines. All three sets of data exhibit a dip around $\tau=0$, but the contrast $\gamma$ is better for a small $\tau_e$.}
\label{Fig:HOMDips}
\end{figure}

In Fig.\ref{Fig:HOMDips}, we present experimental results obtained in the sample presented in Fig.\ref{Fig:HOMSketch}. From average current measurements \cite{Feve2007a}, each of the two sources has been tuned to emit a wavepacket exponentially decaying in time over a time scale $\tau_e=30,100,\SI{180}{\pico\second}$, and the intensity correlations are recorded as function of a delay $\tau$ between arrival times. A clear dip is observed around $\tau=0$ for all three curves, signaling the indistinguishability of the impinging wavepackets, with an increasing width consistent with the increasing temporal width of the wavepackets. The data (symbols) can be fitted with $\Delta q(\tau)=1-\gamma e^{-|\tau|/\tau_e}$ (solid lines),and two important points can be raised. First, the contrast of the two-particle interference is not perfect $\Delta q(\tau=0)=1-\gamma$, which indicates that the wavepackets are only partially undistinguishable. Secondly, the fit yields the temporal width $\tau_e$, which is found to be larger than the one expected from measurement of the current, particularly for short wavepackets.

Though insufficient control of the parameters of the source could lead to improperly prepared wavepackets, the differences are here too strong. They subsequently have to be attributed to interaction effects that alter the propagating wavepackets \cite{Levkivskyi2008}. We show in the next two sections how both the degree of coherence and indistinguishability and the temporal shape of the wavepacket are modified by the presence of Coulomb interaction in the edge channels, and how Hong-ou-Mandel interferometry enables to characterize its effects.

\subsection{Fractionalization in 1D chiral edge channels}

\begin{figure}[htb]%
\includegraphics*[width=0.5\textwidth]{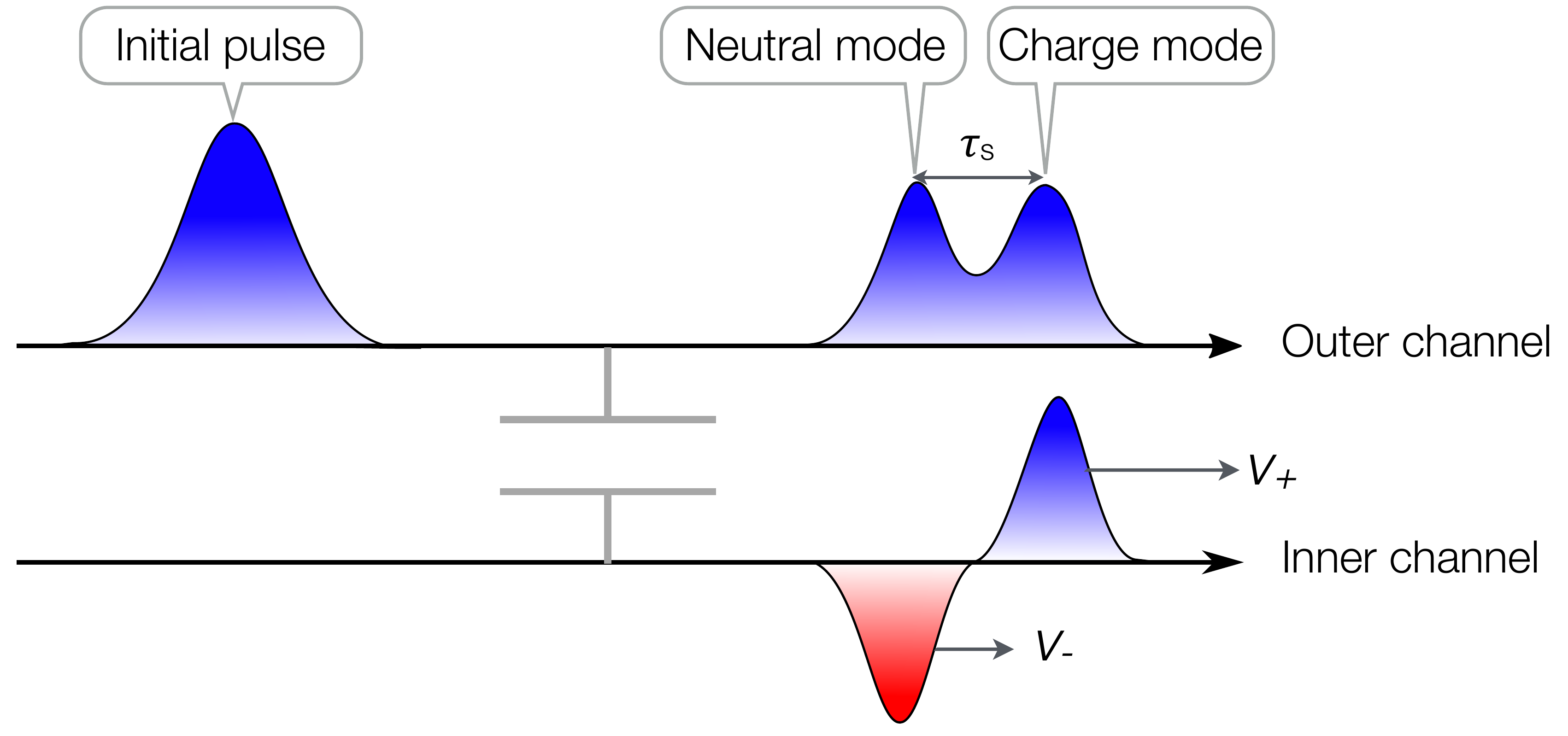}
\caption{Charge fractionalization - A charge density wave created initially in channel 1 can be decomposed in two propagation eigenmodes: a neutral mode $-$, with antisymmetric charge distribution, and a charge mode $+$ with symmetric charge distribution. Coulomb repulsion results in very different velocities, with $v_+\gg v_-$. As propagation takes place, the two modes separate, and the initial charge pulse fractionalizes.
}
\label{Fig:Fractionalization}
\end{figure}

Coulomb interactions within and between edge channels has drastic consequences on the propagation of low-energy wavepackets, that can be conveniently investigated via two-particle interference. In this section, we briefly review the case of filling factor $\nu=2$. In that case, the outer channel is the conductor under study, in which charges are initially injected. It interacts strongly with the inner channel that acts as a well-controlled environment. As both edge channels are strongly coupled, new collective (bosonic) modes appear whose form is known from chiral Luttinger liquid theory. The charge mode corresponds to a symmetric distribution of charge among the two channels and propagates with velocity $v_+$ \cite{Lee1997,Pham2000,Levkivskyi2008,Berg2009,Kamata2010,Kumada2011,Kamata2014}. The neutral mode carries an antisymmetric distribution of charge with velocity $v_-$. Due to Coulomb repulsion, the charge mode is much faster than the neutral mode ($v_+\gg v_-$). A single-electron wavepacket of charge $e$ excited on the outer edge channel splits after propagation on a length $l$ in two charge pulses of charge $e/2$, separated by a time $\tau_s=l/v_-
- l/v_+$, as depicted in Fig.\ref{Fig:Fractionalization}.

\begin{figure*}[htb]%
%\sidecaption
\includegraphics*[width=1\textwidth]{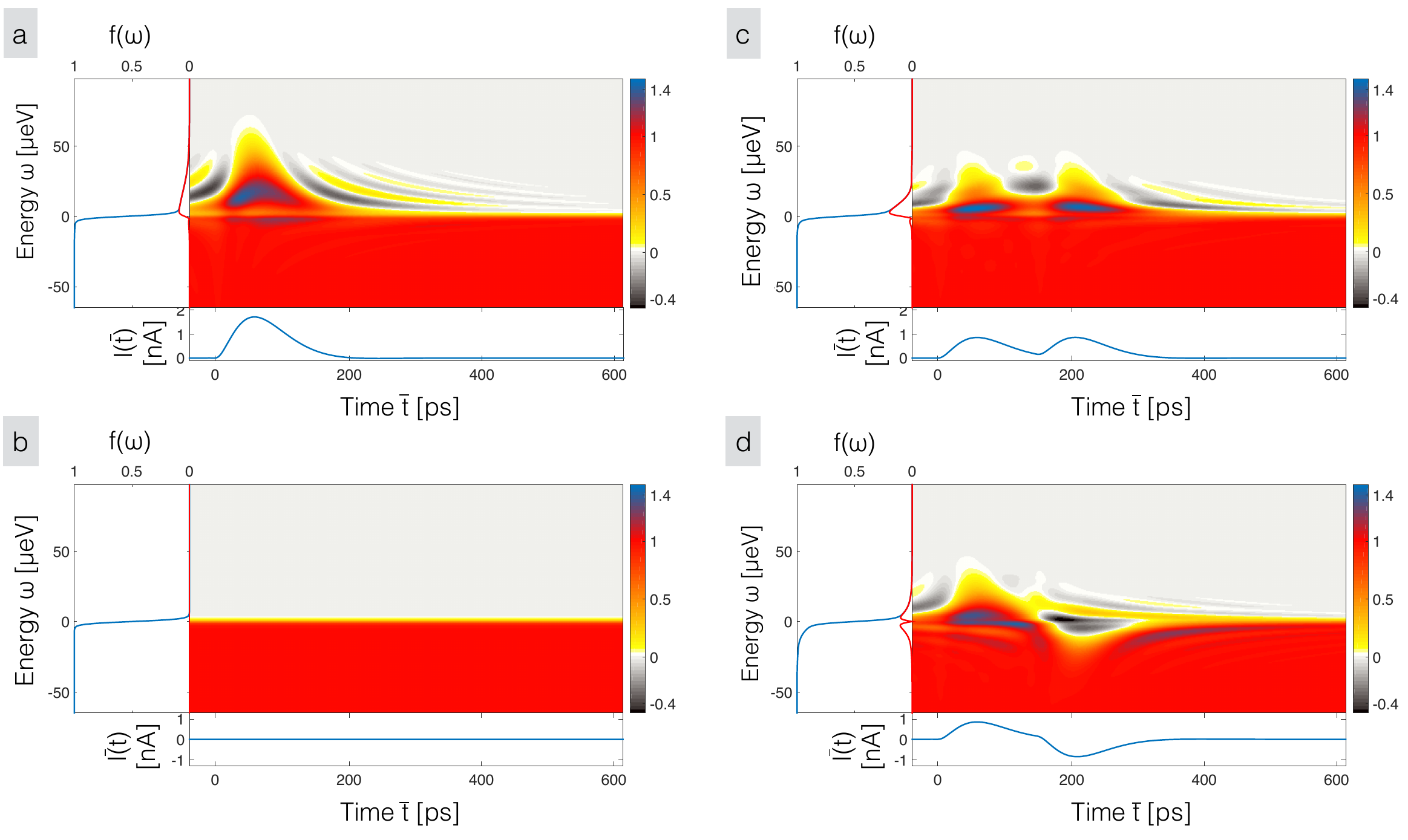}
\caption{Fractionalization of a pulse generated by a voltage driven contact - a,b) Initial Wigner functions $\MWe$, average current $I(\tbar)$ and energy distribution function $f(\omega)$ in the outer (a) and inner (b) channels, at $t=0$. The charge is only present on the outer channel. c,d) Same quantities in the outer (c) and inner (d) channels, for a separation time $\tau_s=\SI{150}{\pico\second}$. The initial charge pulse has split into a fast charge mode and a slow neutral mode. The current exhibits two positive pulses in the outer edge channel, and a dipolar distribution in the inner one. Parameters: $T_{el}=\SI{10}{\milli\kelvin},\tau_s=\SI{150}{\pico\second}$.}
\label{Fig:WignerFractionalization}
\end{figure*}

This many-body problem can not be tackled analytically for arbitrary initial states, but several particular cases have been investigated \cite{Levkivskyi2008,Kovrizhin2011,Kovrizhin2012,Wahl2014,Ferraro2014,Levkivskyi2009,Levkivskyi2012}. The case of a voltage driven contacts is particularly simple, as interactions only appear as a modification of the effective voltage pulse applied on the contact \cite{Freulon2015}. An illustration of the fractionalization process in the Wigner representation is shown for exponentially decaying wavepackets (Fig.\ref{Fig:WignerFractionalization}). The outer channel initially carries all the charge, while the inner channel is empty. After propagation on a length $l$, the pulse has split into the fast charge mode and the slow neutral mode separated by $\tau_s$. As a consequence, the outer edge channel carries a sequence of two pulses with halved amplitude. Moreover, energy relaxation towards the Fermi level is clearly visible between Fig.\ref{Fig:WignerFractionalization}a and Fig.\ref{Fig:WignerFractionalization}c. In contrast, a dipolar charge distribution of electron-hole pairs has been induced in the inner edge channel, as can be seen both in the dipolar nature of the current and in the electron-hole pairs population close to the Fermi level . For more general states, a heavier treatment is required. Numerical simulations \cite{Wahl2014,Ferraro2014} have been performed in the Wigner function framework and shed light on the consequences of interaction on a single electron wavepacket.

In this context, HOM correlations offer a way to access information on the Wigner function of the propagating wavepackets, and thus on their relaxation and decoherence. The next paragraphs detail two different experiments in which interaction effects are particularly prominent.

\subsection{Temporal investigation of single-electron fractionalization}

\begin{figure*}[htb]%
  \includegraphics*[width=\textwidth]{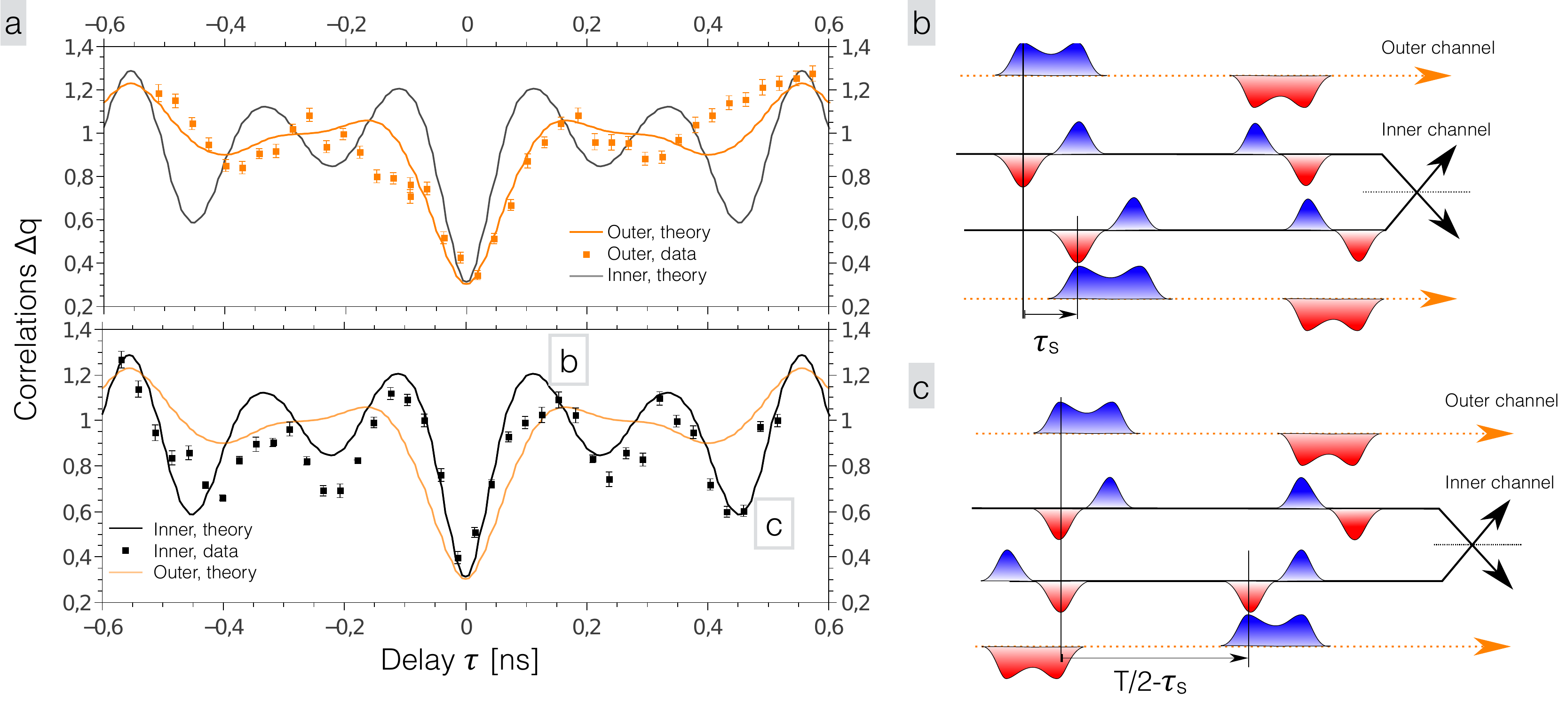}%
  \caption[]{Fractionalization in HOM correlations - a) Correlations $\Delta q_{\rm i/o}$ as a function of delay $\tau$, for the outer (orange, upper panel) and inner (black, lower panel) edge channels. Both channels exhibit non-zero  noise, indicating that some current has been induced in the inner edge channel via capacitive coupling to the outer edge channel. Besides the dip around $\tau=0$, some additional structure is visible with extra peaks and dips. It results from the fractionalization of the charge pulse initially injected on the outer channel. For example, the signal at $\tau\simeq\tau_S$ and $\tau\simeq T/2-\tau_s$ (respectively labeled b and c) are detailed in the other panels. b) As a result of fractionalization, for $\tau\simeq \tau_s$, one expects the pulses of same charges with non-zero overlap to interfere on the outer edge channel, so that $\Delta q_{\rm o}< 1$. In contrast pulses of opposite charges (with little overlap) interfere on the inner edge channel, with $\Delta q_{\rm i}\simeq 1$.
c) On the opposite, for $\tau\simeq T/2-\tau_s$, charges of same signs (respectively opposite signs) reside in the two inner (resp. outer) edge channels, with consequently $\Delta q_{\rm i}<1$ (resp. $\Delta q_{\rm o}\simeq 1$).  }
\label{Fig:HOMFractionalization}
\end{figure*}

The splitting of a charge pulse due to interchannel interaction can be directly probed in the time domain using two-particle interferometry. Indeed, the dependence of $\Delta q(\tau)$ on the time delay $\tau$ encodes the temporal profile of the incoming state. Experimental resolution on $\tau$ can reach a few picoseconds, giving access to time scales shorter than those accessed by time-resolved measurements (typically limited to a few hundreds of picoseconds).

Fig.\ref{Fig:HOMDips} shows that exponentially decaying wavepackets, expected for this type of single electron source, yield exponentially varying correlations $\Delta q$, within experimental resolution. However, as already pointed out, the width seems slightly larger than the one estimated from average current measurements. This is a consequence of the aforementioned fractionalization of the wavepacket. It is particularly prominent on very short wavepackets. Hong-Ou-Mandel interference provides a powerful probe of the alterations of the wavepackets due to fractionalization, as we show below following reference \cite{Freulon2015}.

On a propagation length $l\simeq \SI{3.1}{\micro\meter}$, fractionalization causes a splitting of current pulses in two components separated by a time separation $\tau_s$ estimated around $\tau_s\simeq\SI{70}{\pico\second}$ from a different study of interactions in a similar sample \cite{Bocquillon2013b}.
To maximize the visibility of the fractionalization phenomenon, the impinging wavepackets are subsequently generated with a very short width $\tau_e\simeq \SI{30}{\pico\second}< \tau_s$.

Measurements in both channels \cite{Freulon2015} are summarized in Fig.\ref{Fig:HOMFractionalization}. Importantly, as current pulses are generated in the outer edge channel, and induced by interactions in the inner edge channel, intensity correlations can be measured in almost equal amounts in inner and outer edge channels (denoted $\Delta q_{\rm i}$ and$\Delta q_{\rm o}$ respectively). Thus one can image the complete current distribution.

First around $\tau=0$, the correlations exhibit a dip in both inner and outer channels due to two-particle interference. However their widths are quite different, and the outer channel dip is roughly twice as large as the inner one (\SI{70}{\pico\second} against \SI{40}{\pico\second}). indeed, for $\tau\simeq\pm\tau_s$, a weak anti-bunching effect remains visible in the outer edge channel, $\Delta q_{\rm i}<1$. The increased width reflects the fractionalization that widens the current pulse in the outer edge channel into two pulses of same sign (see Fig.\ref{Fig:Fractionalization}). On the opposite, the correlations overshoot over unity in the outer channel, $\Delta q_o\gtrsim1$. As detailed in \cite{Jonckheere2012,Wahl2014}, this subtle effect arises from overlap at finite temperature between an electron-like and a hole-like current pulse. This confirms the dipolar nature of the current distribution flowing in the inner edge channel, in contrast with the monotonous trend observed in the outer edge channel.
Further signatures of fractionalization can be obtained from measurements for larger delays $\tau=\pm T/2$, when emissions of electron-like wavepacket on source 1 is synchronized with hole-like wavepackets in source 2, and vice-versa. Both channels exhibit a weak bunching effect from thermal overlap, with $\Delta q_{\rm i/o}(\pm T/2)\gtrsim 1$. In analogy with measurements around $\tau=0$, the peak in $\Delta q_{\rm o}$ is twice as large as in $\Delta q_{\rm i}$, that rapidly drops below 1 around $\tau\simeq T/2-\tau_s$.

One can model the fractionalization of the wavepackets emitted in each source in the Wigner representation as represented in Fig.\ref{Fig:WignerFractionalization}, and their overlap to obtain $\Delta q_{\rm i/o}(\tau)$. The results are shown as plain lines in Fig.\ref{Fig:HOMFractionalization}. A good agreement can be found with experimental data, if one takes into account the imperfections of the rf excitation drive (finite rise time and finite number of harmonics).

\subsection{Visibility and interaction}

\begin{figure}[htb]%
\includegraphics*[width=0.48\textwidth]{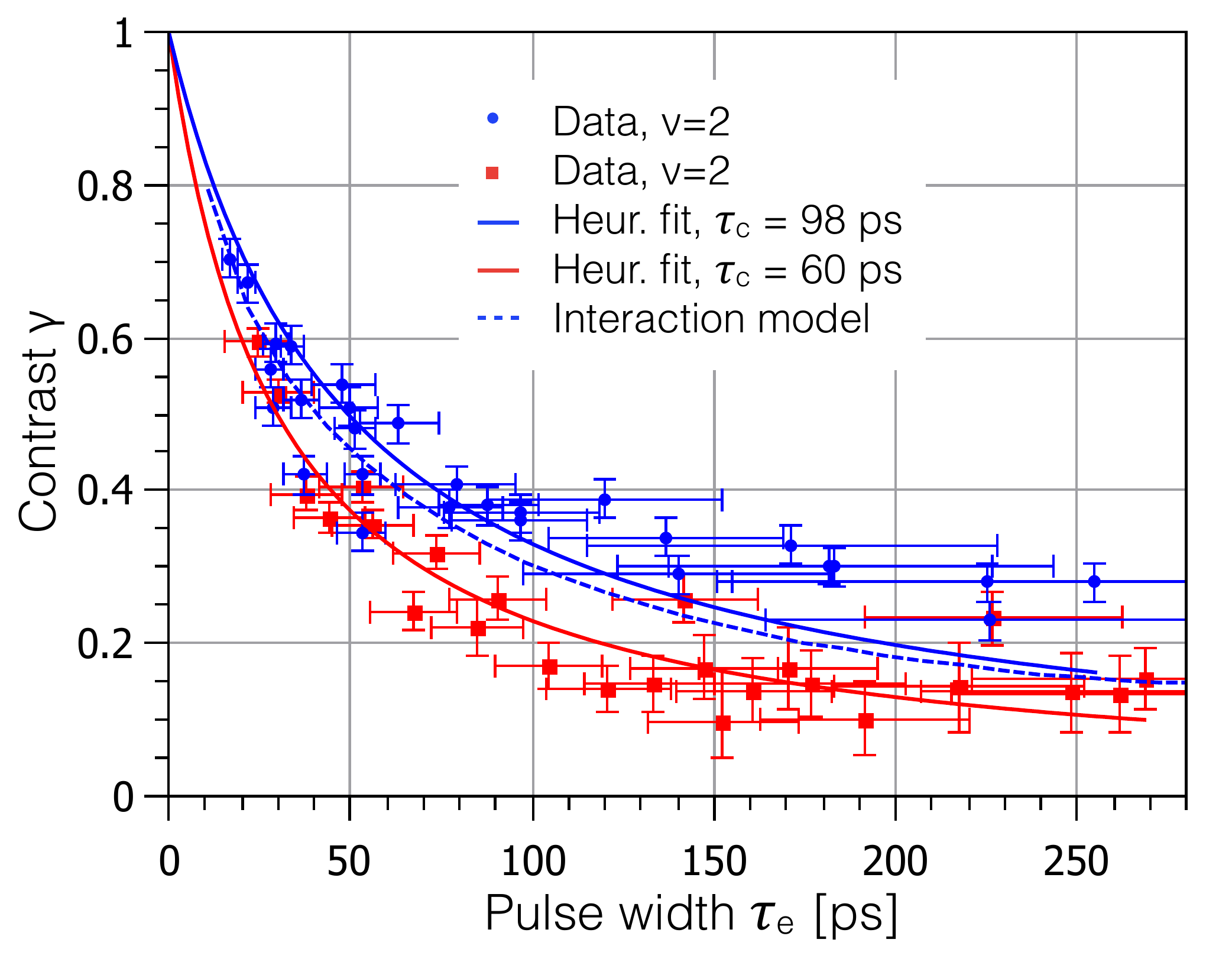}
\caption{Contrast of HOM correlations - The contrast $\gamma$ (extracted from exponential fits) is plotted as a function of the temporal width $\tau$, for filling factors $\nu=2$ and $\nu=3$. It confirms that $\gamma$ decreases for larger $\tau_e$. As solid lines, heuristic fits $\gamma(\tau_e)=(1+\frac{2\tau_e}{\tau_c})^{-1}$ show a good agreement and yield the coherence time $\tau_c$. The blue dashed line presents the contrast computed from an interacting model.
}
\label{Fig:HOMContrasts}
\end{figure}

Beyond the deformation of the shape of incoming states in time domain, the interaction process also leads to the relaxation and decoherence of elementary electronic excitations. The contrast of the two-particle Hong-Ou-Mandel effect observed at $\tau=0$ provides a measurement of the degree of coherence and subsequently enables to define a phenomenological coherence time $\tau_c$ for the incoming wavepackets \cite{Marguerite2016}. Naively, one expects that short wavepackets for which $\tau_e<\tau_c$ are less sensitive to decoherence than longer ones ($\tau_e>\tau_c$) which require to preserve the wave packet phase coherence on much longer times. Phenomenologically, we assume that interactions affect the non-diagonal parts of the coherence function such that $\Delta\mathcal{G}^{(e)}(t, t')\to e^{-|t-t'|/\tau_c}\, \Delta\mathcal{G}^{(e)}(t, t')$. Then, only time components $(t, t')$ with $|t-t'| \leq \tau_c$ of the wavepacket can interfere on the splitter whereas components for $|t-t'| \geq \tau_c$ are subject to random partitioning. It can then be shown that for exponential wavepackets, the contrast simply reads $\gamma=(1+\frac{2\tau_e}{\tau_c})^{-1}$.
Fig.\ref{Fig:HOMContrasts} shows measurements of $\gamma$ recorded for values of $\tau_e$ ranging between 20 and \SI{250}{\pico\second}, for filling factors $\nu=2$ and $\nu=3$. The agreement with the previous phenomenological model (shown as plain lines) is good, for fitting parameters of $\tau_c=\SI{98}{\pico\second}$ ($\nu=2$) and $\tau_c=\SI{60}{\pico\second}$ ($\nu=3$). The coherence time $\tau_c$ depends on the filling factor $\nu$, confirming that decoherence occurs mostly during propagation. For increasing filling factor, the number of edge channel increases, and they move closer to one another. The observed decrease of the coherence time $\tau_c$ suggests that capacitive inter-channel is prominent, as already evidenced in several experiments \cite{Neder2006,LeSueur2010,Altimiras2010,Bocquillon2013b,Inoue2013}.  As already mentioned, a more exact treatment of interactions can be obtained via bosonization techniques and gives similar predictions for the contrast $\gamma$ \cite{Wahl2014} (see blue dashed line on Fig.\ref{Fig:HOMContrasts}).

The two-particle interference between two supposedly identical wavepackets measures their degree of similarity, and thus reveals signatures of their coherence and temporal shape. In the last section of this article, we turn to the case of an unknown source, that is being progressively characterized by measuring its overlap with reference sources. This enables to obtain first a spectroscopy of the wavepacket in the energy domain, and more generally yields a tomography protocol to reconstruct $\Delta\MWe$.

\section{Spectroscopy and tomography of single-electron wavepackets}
\label{Section:Tomo}

The main idea of this section is to use Eq.\ref{Eq:QbarHOM} with an unknown source in input 1 and reference sources in 2, to be able to access $\MWe_1$ from its overlap with reference values of $\MWe_2$. We divide this section in three parts. First, we study the case of an unbiased Fermi sea in arm 2, in which source 1 only interferes with thermal excitations of the sea. Second, we discuss the case of interference with various reference sources as spectroscopy and tomography protocols.

\subsection{Two-particle interference with thermal excitations}
\label{Section:Tomo:HBT}

\begin{figure}[htb]%
  \includegraphics*[width=.48\textwidth]{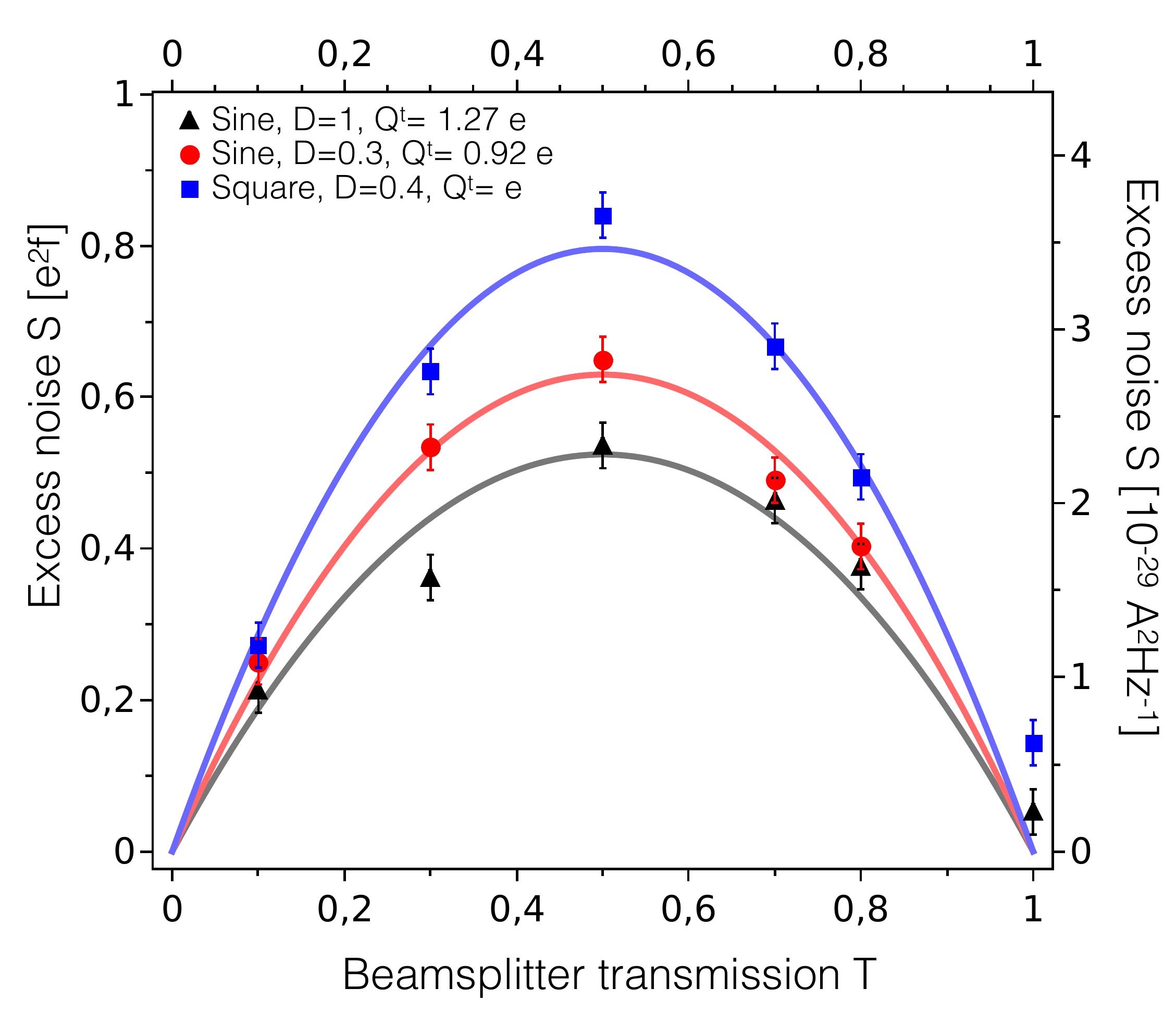}%
  \caption[]{Excess auto-correlations $\Delta \Sbar_{33}$ a function of the beamsplitter transmission $\mathcal{T}$ in the HBT configuration : source 1 is fed with a single electron source, while source 2 is a Fermi sea. As indicated in the legend, three sets of parameters are studied (varying transmission $D$ of the quantum dot QPC, and using a sine or square drive). The dependence $\Delta \Sbar\propto \mathcal{T}(1-\mathcal{T})$ is observed in all three cases, but the maximum value of $\Delta\Sbar$ strongly varies.}
    \label{Fig:HBTvsD}
\end{figure}

In this part, we assume that input 2 is an unbiased Fermi sea, at finite temperature $T_{el}$ such that
\begin{eqnarray}
\Qbar&=&\Qbar_{\rm HBT,1}\nonumber\\
&=&2 e^2\int \frac{d\omega}{2\pi}\,\overline{\Delta\MWe_{1}(\tbar,\omega)}^{\tbar}\left(1-2f_{0}(\omega)\right)
\end{eqnarray}
As defined in Eq.\ref{Eq:DefDeltaF}, $\overline{\Delta\MWe_{1}(\tbar,\omega)}^{\tbar}=\Delta f_1(\omega)$ is the excess electronic distribution function of electrons and holes (with respect to the Fermi sea $f_0(\omega)$) produced by source 1 \cite{Bocquillon2012}. At $T_{el}=0$, $\Qbar$ is directly proportional to the total number of incoming excess excitations $\Delta N_1$ (electrons and holes) emitted by source 1 in time $T$:  
 \begin{eqnarray}
& &\Delta N_1  =  T\left[\int_{0}^{+\infty} \frac{d\omega}{2\pi} \Delta f_1(\omega) - \int_{-\infty}^{0} \frac{d\omega}{2\pi} \Delta f_1(\omega)\right] \\
& & \Qbar_{T_{el}=0}  = \frac{2 e^2}{T}  \Delta N_1
\end{eqnarray}
This expression simply represents the random partitioning of $\Delta N_1$ classical particles reaching the splitter in time $T$.
At finite temperature, the term $1-2f_{0}(\omega)$ suppresses the contribution of electron and holes in the energy range $k_BT_{el}$ around the Fermi energy.  This suppression can be interpreted as a noise reduction coming from two-particle interferences between the excitations emitted by source 1 at energy $\omega$, and thermal excitations at the same energy in input 2. It can in particular be recast as :
\begin{eqnarray}
\Qbar_{T_{el}\neq0}  &=&  \Qbar_{T_{el}=0} \nonumber\\
&&-4e^2\int \frac{d\omega}{2\pi}\, \overline{\Delta\MWe_{1}(\tbar,\omega)\Delta\MWe_{2}(\tbar,\omega)}^{\tbar}\\
\Delta\MWe_{2}(\omega)&=&f_{0,T_{el}\neq0}(\omega)-f_{0,T_{el}=0}(\omega)
\end{eqnarray}
$\Delta\MWe_{2}$ then represents the excess Wigner function of thermal excitations, created around the Fermi level of a sea at zero temperature.

The noise reduction can be quite drastic if source 1 generates excitations close to the Fermi level, where most thermal excitations reside. The two-particle interference with thermal excitations thus provides a first probe of the energy distribution $\Delta f_1$ by measuring its overlap with the $\Delta\MWe_2$. In ref.\cite{Bocquillon2012}, three types of wavepackets are generated by exciting the mesoscopic capacitor either with a square or a sinusoidal signal, and by changing its coupling to the quantum Hall edge channel (transmission $D$). All three curves show the same expected dependence on the transmission $\Delta \Sbar\propto \mathcal{T}(1-\mathcal{T})$. However, while the incoming charge per period (indicated by $Q^t$) is slightly larger for the sine wave excitation, the amplitude of the measured partition noise is much smaller in comparison to the case of a square drive. It reflects the fact that a square drive creates energy-resolved packets flying well above the Fermi sea, while a sine drive generates excitations close to the Fermi level, more likely to interfere with thermal excitations. Similarly, decreasing the coupling (transmission $D$) increases the dwell time of electrons in the dot, that are then emitted at higher energies.
Strong temperature effects can also be observed when measuring partition noise for Levitons \cite{Dubois2013b}, that are very sensitive to finite-temperature effects as they live close to the Fermi level.

This section shows how two-particle interference can be used to access partial information on the energy distribution $\Delta f_1(\omega)$ of incoming wavepackets by analyzing the two-particle interference with thermal excitations, that reduces the amplitude of the shot noise. Not only using thermal excitations, but a carefully engineered reference state in arm 2, we now show how one can reconstruct $\Delta f_1(\omega)$ and even $\Delta\MWe_{1}(\tbar,\omega)$ as a whole.

\subsection{Spectroscopy and tomography of single-electron wavepackets}

Probing an unknown Wigner function $\Delta \MWe_{1}(\tbar,\omega)$ with a thermal source suffers from two major limitations. Firstly, there is no possibility to vary the energy scanned by the thermal source such that, even if useful information can be extracted, the full energy distribution cannot be reconstructed. Secondly, as a thermal source is stationary, no information can be obtained on the dynamics of the source encoded in the time dependence of $\Delta \MWe_{1}(\tbar,\omega)$. Getting access to temporal information requires to use an a.c. source as a probe.  As we will see, a combination of a d.c. bias and a small amplitude a.c. sinusoidal drive can be used to fully reconstruct an unknown Wigner distribution $\Delta \MWe_{1}(\tbar,\omega)$ \cite{Grenier2011NJP}. For a $T=2\pi/\Omega$-periodic source, $\Delta \MWe_{1}(\tbar,\omega)$ can be written in Fourier representation:

\begin{eqnarray}
\Delta\MWe_1(\tbar,\omega)=\sum_{n=-\infty}^{+\infty} \Delta\MWe_{1,n}(\omega)e^{-in\Omega\tbar}
\end{eqnarray}

\begin{figure}[htb]%

  \includegraphics*[width=.45\textwidth]{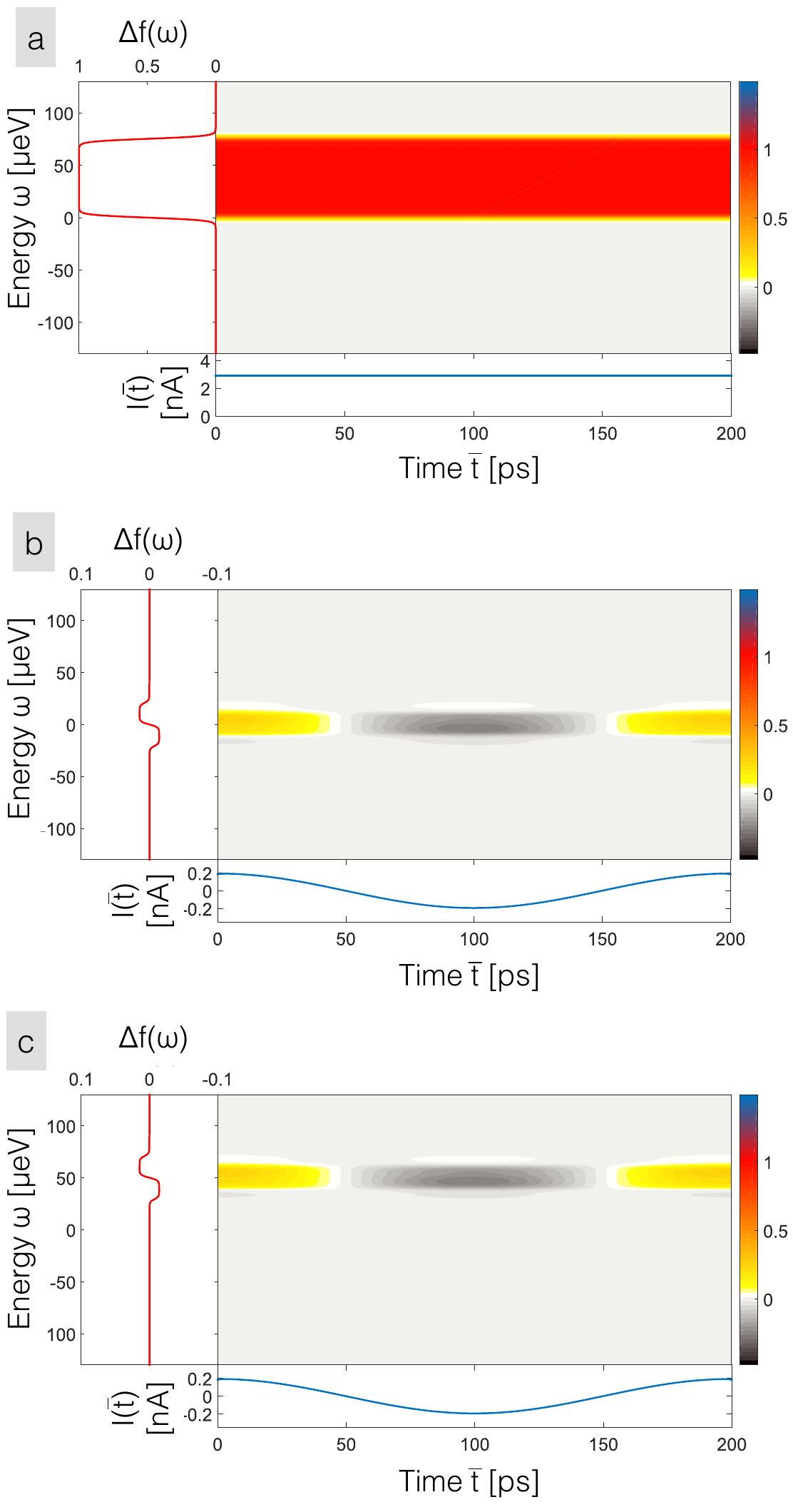}%
  \caption[]{Excess Wigner functions $\Delta\MWe$, energy distribution function $\Delta f(\omega)$ and current $I(\tbar)$ of the injected reference signal - a) Pure dc-bias for $\Vdc= \SI{75}{\micro\volt}, T_{el}=\SI{15}{\milli\kelvin}$. b) Pure ac-bias, with $T=\SI{200}{\pico\second}, \Vac=\SI{5}{\micro\volt}, T_{el}=\SI{15}{\milli\kelvin},n=1,\phi=0$. c) Mixed ac and dc bias, with $e\Vdc=\SI{-50}{\micro\volt}, T=\SI{200}{\pico\second}, e\Vac=\SI{5}{\micro\volt}, n=1,\phi=0$. Please note that in this case, the excess Wigner function is defined with respect to a Fermi sea at $\mu=-e\Vdc$ and not $\mu=0$. 
  }
    \label{Fig:WignerRef}
\end{figure}

We can first focus on the $n=0$ term, $\Delta\MWe_{1,0}(\omega)$,  which is nothing but the excess energy distribution $\Delta f_1(\omega)$. By applying a d.c. bias $\mu=-eV_{dc}$ on source 2, the excess Wigner function in 2 can be written as $\Delta\MWe_{2}(\omega)=f_0(\omega-\omega_{dc})-f_0(\omega)$ (with $\omega_{dc}=-e\Vdc/\hbar)$, which simply is a rectangular function for the energy window $[0,-e\Vdc]$ (assuming $\Vdc \leq 0)$ with thermal smearing (see Fig.\ref{Fig:WignerRef}a).

Consequently, an electron emitted in arm 1 at energy $\hbar\omega<-e\Vdc$ impinges on the beamsplitter together with an electron from the biased Fermi sea in input 2, so that HOM two-particle antibunching effect will then occur in this window. Tuning the energy window via the applied dc bias $\Vdc$, one modulates this HOM interference which is sufficient to compute the excess energy distribution $\Delta f_1(\omega)$ in the unknown input arm 1. This is more explicitly evidenced by writing the HOM contribution to the noise in this case:
\begin{eqnarray}
&&\Qbar_{\rm HOM}=\nonumber\\
\qquad&&-4e^2\int \frac{d\omega}{2\pi}\, \Delta f_1(\omega) \big[f_0(\omega-\omega_{dc})-f_0(\omega)\big]
\end{eqnarray}
By measuring the derivative of the HOM noise with respect to the dc bias, one measures $\Delta f_1(\omega)$ at $\omega = \omega_{dc}$ convoluted by the thermal smearing on the energy window $k_BT$:
\begin{eqnarray}
-\frac{\partial \Qbar_{\rm HOM}}{\partial V_{dc}} & =& \frac{4 e^3}{h} \int d\omega \Delta f_1(\omega)  \frac{-\partial f_{0}}{\partial \omega}(\omega-\omega_{dc}) 
\end{eqnarray}
If the temperature is well known, it is possible to reconstruct the true energy distribution $\Delta f_1(\omega)$ using Wiener deconvolution methods \cite{Wiener1949}.

As mentioned above, accessing the dependence on $\tbar$ of the unknown Wigner function in 1 requires not only selectivity on the $\omega$ axis, but also non-stationary references sources. In order to pick the $n^{\rm th}$ component of $\Delta\MWe_1(\tbar,\omega)$, one needs a probe on input 2 which Wigner distribution depends sinusoidally on time at frequency $n\Omega$. This the exact dependence\footnote{In this particular case, the excess Wigner function is defined with respect to a Fermi sea at $\mu=-e\Vdc$ and not $\mu=0$.} one obtains in the case where a small amplitude $eV_{ac} \ll n\hbar\Omega$ sinusoidal drive is applied on input 2 \cite{Ferraro2013}:
\begin{eqnarray}
\Delta\MWe_2(\tbar,\omega)& =& \frac{eV_{ac}}{\hbar} \cos{(n\Omega t + \phi)} g_n(\omega -\omega_{dc}) \\
g_n(\omega) & =& \frac{ f_{0}(\omega-n\Omega/2) - f_{0}(\omega+n\Omega/2) }{n\Omega}
\end{eqnarray}
Examples of probe Wigner functions (for $n=1$ and $\Vdc=0$ and \SI{-50}{\micro\volt}) are presented in Fig.\ref{Fig:WignerRef}b and \ref{Fig:WignerRef}c. It has $n+1=2$ nodal lines and occupies an energy range defined by the energy window of $g_n(\omega)$ which has a width $n\Omega$ centered on $\omega_{dc}$. The procedure is then straightforward, time information is accessed by varying the phase $\phi$  and frequency $n\Omega$ of the a.c. drive while energy information is obtained by varying the d.c. bias. Real and imaginary parts of $\Delta \MWe_{1,n}$ can then be obtained by measuring the HOM contribution to the noise, and its dependence with phase $\phi$ of the probe:

\begin{eqnarray}
& & \frac{\Qbar_{\rm HOM}^{\phi=0} -\Qbar_{\rm HOM}^{\phi=\pi}}{V_{ac}}  = \nonumber \\ 
 & & \qquad \qquad - \frac{4e^3}{h} \int d\omega \, {\rm Re} \big[\Delta\MWe_n(\omega)\big] g_n(\omega-\omega_{dc}) \\
&&\frac{\Qbar_{\rm HOM}^{\phi=\pi/2} -\Qbar_{\rm HOM}^{\phi=3\pi/2}}{V_{ac}}   =  \nonumber\\
 & & \qquad \qquad \frac{4e^3}{h} \int d\omega \, {\rm Im} \big[\Delta\MWe_n(\omega)\big] g_n(\omega-\omega_{dc})
\end{eqnarray}

Here also, the exact real and imaginary parts of $\Delta\MWe_n(\omega)$ are convoluted with the function $g_n$ and deconvolution techniques are required in order to reconstruct the exact $\Delta\MWe_{1,n}(\omega)$.

To implement such protocols, most difficulties reside in the measurement of low levels of HOM correlations in combination with the use of multiple rf excitations signals. However, this protocol can be simplified if some assumptions are made a priori on the unknown state. For example, it is experimentally challenging to access a large number of harmonics $\Delta\MWe_{1,n}$, and it is easier to describe states for which $|\Delta W_{1,n}|$ in general decays rapidly with harmonic $n$. Besides, the Wiener deconvolution process requires an accurate measurement of the dependence of $\Delta W_{1,n}$ on $\omega_{\rm dc}$.  States created by a voltage drive directly on an ohmic contact are however simpler to characterize, as they are fully parameterized by a discrete set of numbers. Indeed, in the framework of photon-assisted shot noise, the state created in the conductor is only given by the photon-assisted transition amplitudes $p_n$ \cite{Dubois2013}, with:
\begin{eqnarray}
p_n&=&\frac{1}{T}\int_0^T dt e^{-\frac{ie}{\hbar}\int_{-\infty}^tV(t')dt'}e^{-in\Omega t}\\
\MWe(\tbar,\omega)&=&\sum_{n,m=-\infty}^\infty p_n p^*_m e^{i(m-n)\Omega\tbar}\nonumber\\
&&\quad\quad f_\mu(\omega-\frac{\Omega}{2}(n+m))
\end{eqnarray}
Under these assumptions, Jullien and co-workers have been able to experimentally implement this scheme \cite{Jullien2014}, by analysing the HOM-like noise in this shot noise framework. They have thus realized the first tomography protocol of voltage pulses and obtained a fairly accurate description of the Wigner function of the state created by a train of Lorentzian pulses (Levitons). The complete tomography of arbitrary states, such as states generated by a time-dependent scatterer, is however yet to be implemented.

\section{Conclusions}

As discussed in this review, the mere measurement of the time-dependent electrical current does not allow for a non-ambiguous characterization of the electron states propagating in a ballistic conductor.
Indeed, when studying time-dependent transport, it is crucial to capture both the energetic and temporal aspects of the propagating states. In this context, a mixed time-frequency representation such as the Wigner function is relevant as it encompasses all the single particle properties of the system. In particular, this theoretical tool is particularly well suited to the case of single particle states at the heart of the present Focus Issue.

In this review, we have discussed the use of two-particle interference effect to probe, chracterize or even reconstruct the Wigner function of single electron states propagating in ballistic conductors. Single-particle coherence is in principle more easily accessed in conventional one-particle interferometers (Mach-Zehnder interferometer for example), but suffer from decoherence effects within the interferometer. In this context, two-particle interferometer offer a powerful alternative and allow for a punctual decoherence-free measurement of single particle coherence. Though two-particle interferometry has only been recently implemented for electrons, several recent studies illustrate the richness and versatility of this method.

Future developments can already be envisioned. First, the tools of electron quantum optics could be adapted to other ballistic conductors, for example to investigate excitations in topological matter \cite{Hofer2013,Inhofer2013,Ferraro2014a,Ferraro2014b,Calzona2015,Dolcetto2016}. Another route consists in extending the previously introduced tools, in order to go beyond the single-particle picture and capture correlations and entanglement \cite{Splettstoesser2009,Moskalets2014,Thibierge2016}. Such efforts contribute to the development of quantum signal processing based on electron quantum optics \cite{Roussel2016}.

\begin{acknowledgement}
We warmly thank Janine Splettstoesser and Rolf Haug for setting up this special issue. This work has been supported by ANR grants '1-shot' (ANR-2010-BLANC-0412) and '1-shot:reloaded' (ANR-14-CE32-0017), and ERC Consolidator grant 'EQuO' (No. 648236).
\end{acknowledgement}

\bibliographystyle{pss}
\bibliography{BibPSS}

\providecommand{\WileyBibTextsc}{}
\let\textsc\WileyBibTextsc
\providecommand{\othercit}{}
\providecommand{\jr}[1]{#1}
\providecommand{\etal}{~et~al.}


\begin{thebibliography}{[10]}

\bibitem{Moskalets2002}% article
 \textsc{M.~Moskalets} and  \textsc{M.~B{\"{u}}ttiker}\iffalse {Floquet
  scattering theory of quantum pumps}\fi,
 \jr{Physical Review B} \textbf{66}, 205320 (2002).


\bibitem{Moskalets2008}% article
 \textsc{M.~Moskalets},  \textsc{P.~Samuelsson},  and
  \textsc{M.~B{\"{u}}ttiker}\iffalse {Quantized Dynamics of a Coherent
  Capacitor}\fi,
 \jr{Physical Review Letters} \textbf{100}, 086601 (2008).


\bibitem{Henny1999}% article
 \textsc{M.~Henny},  \textsc{S.~Oberholzer},  \textsc{C.~Strunk},
  \textsc{T.~Heinzel},  \textsc{K.~Ensslin},  \textsc{M.~Holland},  and
  \textsc{C.~Sch{\"{o}}nenberger}\iffalse {The Fermionic Hanbury Brown and
  Twiss Experiment}\fi,
 \jr{Science} \textbf{284}, 296 (1999).


\bibitem{Oliver1999}% article
 \textsc{W.~Oliver},  \textsc{J.~Kim},  \textsc{R.~Liu},  and
  \textsc{Y.~Yamamoto}\iffalse {Hanbury Brown and Twiss-type experiment with
  electrons}\fi,
 \jr{Science} \textbf{284}, 299--301 (1999).


\bibitem{Ji2003}% article
 \textsc{Y.~Ji},  \textsc{Y.~Chung},  \textsc{D.~Sprinzak},
  \textsc{M.~Heiblum},  \textsc{D.~Mahalu},  and  \textsc{H.~Shtrikman}\iffalse
  {An Electronic Mach-Zehnder Interferometer}\fi,
 \jr{Nature} \textbf{422}, 415--8 (2003).


\bibitem{Samuelsson2004}% article
 \textsc{P.~Samuelsson},  \textsc{E.~Sukhorukov},  and
  \textsc{M.~B{\"{u}}ttiker}\iffalse {Two-Particle Aharonov-Bohm Effect and
  Entanglement in the Electronic Hanbury Brown–Twiss Setup}\fi,
 \jr{Physical Review Letters} \textbf{92}, 026805 (2004).


\bibitem{Splettstoesser2009}% article
 \textsc{J.~Splettstoesser},  \textsc{M.~Moskalets},  and
  \textsc{M.~B{\"{u}}ttiker}\iffalse {Two-Particle Nonlocal Aharonov-Bohm
  Effect from Two Single-Particle Emitters}\fi,
 \jr{Physical Review Letters} \textbf{103}, 076804 (2009).


\bibitem{Neder2006}% article
 \textsc{I.~Neder},  \textsc{M.~Heiblum},  \textsc{Y.~Levinson},
  \textsc{D.~Mahalu},  and  \textsc{V.~Umansky}\iffalse {Unexpected Behavior in
  a Two-Path Electron Interferometer}\fi,
 \jr{Physical Review Letters} \textbf{96}, 016804 (2006).


\bibitem{Roulleau2007}% article
 \textsc{P.~Roulleau},  \textsc{F.~Portier},  \textsc{D.\,C. Glattli},
  \textsc{P.~Roche},  \textsc{A.~Cavanna},  \textsc{G.~Faini},
  \textsc{U.~Gennser},  and  \textsc{D.~Mailly}\iffalse {Finite bias visibility
  of the electronic Mach-Zehnder interferometer}\fi,
 \jr{Physical Review B} \textbf{76}, 161309 (2007).


\bibitem{Levkivskyi2008}% article
 \textsc{I.~Levkivskyi} and  \textsc{E.~Sukhorukov}\iffalse {Dephasing in the
  electronic Mach-Zehnder interferometer at filling factor $\nu$=2}\fi,
 \jr{Physical Review B} \textbf{78}, 045322 (2008).


\bibitem{Buttiker1993}% article
 \textsc{M.~B{\"{u}}ttiker},  \textsc{H.~Thomas},  and
  \textsc{A.~Pr{\^{e}}tre}\iffalse {Mesoscopic capacitors}\fi,
 \jr{Physics Letters A} \textbf{180}, 364--369 (1993).


\bibitem{Gabelli2006a}% article
 \textsc{J.~Gabelli},  \textsc{G.~F{\`{e}}ve},  \textsc{J.\,M. Berroir},
  \textsc{B.~Pla{\c{c}}ais},  \textsc{A.~Cavanna},  \textsc{B.~Etienne},
  \textsc{Y.~Jin},  and  \textsc{D.~Glattli}\iffalse {Violation of Kirchhoff's
  laws for a coherent RC circuit.}\fi,
 \jr{Science} \textbf{313}, 499--502 (2006).


\bibitem{Feve2007a}% article
 \textsc{G.~F{\`{e}}ve},  \textsc{A.~Mah{\'{e}}},  \textsc{J.\,M. Berroir},
  \textsc{T.~Kontos},  \textsc{B.~Pla{\c{c}}ais},  \textsc{D.~Glattli},
  \textsc{A.~Cavanna},  \textsc{B.~Etienne},  and  \textsc{Y.~Jin}\iffalse {An
  on-demand coherent single-electron source}\fi,
 \jr{Science} \textbf{316}, 1169 (2007).


\bibitem{Gabelli2012}% article
 \textsc{J.~Gabelli},  \textsc{G.~F{\`{e}}ve},  \textsc{J.\,M. Berroir},  and
  \textsc{B.~Pla{\c{c}}ais}\iffalse {A coherent RC circuit.}\fi,
 \jr{Reports on progress in physics.} \textbf{75}, 126504 (2012).


\bibitem{Fricke2011}% article
 \textsc{L.~Fricke},  \textsc{F.~Hohls},  \textsc{N.~Ubbelohde},
  \textsc{B.~Kaestner},  \textsc{V.~Kashcheyevs},  \textsc{C.~Leicht},
  \textsc{P.~Mirovsky},  \textsc{K.~Pierz},  \textsc{H.~Schumacher},  and
  \textsc{R.~Haug}\iffalse {Quantized current source with mesoscopic
  feedback}\fi,
 \jr{Physical Review B} \textbf{83}, 193306 (2011).


\bibitem{Mirovsky2013}% article
 \textsc{P.~Mirovsky},  \textsc{L.~Fricke},  \textsc{F.~Hohls},
  \textsc{B.~Kaestner},  \textsc{Ch.Leicht},  \textsc{K.~Pierz},
  \textsc{J.~Melcher},  and  \textsc{H.~Schumacher}\iffalse {Towards quantized
  current arbitrary waveform synthesis}\fi,
 \jr{Journal of Applied Physics} \textbf{113}, 213704 (2013).


\bibitem{Fletcher2013}% article
 \textsc{J.~Fletcher},  \textsc{P.~See},  \textsc{H.~Howe},
  \textsc{M.~Pepper},  \textsc{S.~Giblin},  \textsc{J.~Griffiths},
  \textsc{G.~Jones},  \textsc{I.~Farrer},  \textsc{D.~Ritchie},
  \textsc{T.~Janssen},  and  \textsc{M.~Kataoka}\iffalse {Clock-Controlled
  Emission of Single-Electron Wave Packets in a Solid-State Circuit}\fi,
 \jr{Physical Review Letters} \textbf{111}, 216807 (2013).


\bibitem{Ubbelohde2014}% article
 \textsc{N.~Ubbelohde},  \textsc{F.~Hohls},  \textsc{V.~Kashcheyevs},
  \textsc{T.~Wagner},  \textsc{L.~Fricke},  \textsc{B.~K{\"{a}}stner},
  \textsc{K.~Pierz},  \textsc{H.~Schumacher},  and  \textsc{R.~Haug}\iffalse
  {Partitioning of on-demand electron pairs}\fi,
 \jr{Nature Nanotechnology} \textbf{10}, 46--49 (2014).


\bibitem{Waldie2015}% article
 \textsc{J.~Waldie},  \textsc{P.~See},  \textsc{V.~Kashcheyevs},
  \textsc{J.~Griffiths},  \textsc{I.~Farrer},  \textsc{G.~Jones},
  \textsc{D.~Ritchie},  \textsc{T.~Janssen},  and  \textsc{M.~Kataoka}\iffalse
  {Measurement and control of electron wave packets from a single-electron
  source}\fi,
 \jr{Physical Review B} \textbf{92}, 125305 (2015).


\bibitem{Levitov1996}% article
 \textsc{L.~Levitov},  \textsc{H.\,W. Lee},  and  \textsc{G.~Lesovik}\iffalse
  {Electron Counting Statistics and Coherent States of Electric Current}\fi,
 \jr{Journal of Mathematical Physics} \textbf{37}, 4845 (1996).


\bibitem{Ivanov1997}% article
 \textsc{D.~Ivanov},  \textsc{H.~Lee},  and  \textsc{L.~Levitov}\iffalse
  {Coherent states of alternating current}\fi,
 \jr{Physical Review B} \textbf{56}, 6839 (1997).


\bibitem{Dubois2013b}% article
 \textsc{J.~Dubois},  \textsc{T.~Jullien},  \textsc{F.~Portier},
  \textsc{P.~Roche},  \textsc{A.~Cavanna},  \textsc{Y.~Jin},
  \textsc{W.~Wegscheider},  \textsc{P.~Roulleau},  and  \textsc{D.\,C.
  Glattli}\iffalse {Minimal-excitation states for electron quantum optics using
  levitons.}\fi,
 \jr{Nature} \textbf{502}, 659 (2013).


\bibitem{Bocquillon2014}% article
 \textsc{E.~Bocquillon},  \textsc{V.~Freulon},  \textsc{F.~Parmentier},
  \textsc{J.\,M. Berroir},  \textsc{B.~Pla{\c{c}}ais},  \textsc{C.~Wahl},
  \textsc{J.~Rech},  \textsc{T.~Jonckheere},  \textsc{T.~Martin},
  \textsc{C.~Grenier},  \textsc{D.~Ferraro},  \textsc{P.~Degiovanni},  and
  \textsc{G.~F{\`{e}}ve}\iffalse {Electron quantum optics in ballistic chiral
  conductors}\fi,
 \jr{Annalen der Physik} \textbf{526}, 1--30 (2014).


\bibitem{Wahl2014}% article
 \textsc{C.~Wahl},  \textsc{J.~Rech},  \textsc{T.~Jonckheere},  and
  \textsc{T.~Martin}\iffalse {Interactions and Charge Fractionalization in an
  Electronic Hong-Ou-Mandel Interferometer}\fi,
 \jr{Physical Review Letters} \textbf{112}, 046802 (2014).


\bibitem{Ferraro2014}% article
 \textsc{D.~Ferraro},  \textsc{B.~Roussel},  \textsc{C.~Cabart},
  \textsc{E.~Thibierge},  \textsc{G.~F{\`{e}}ve},  \textsc{C.~Grenier},  and
  \textsc{P.~Degiovanni}\iffalse {Real-Time Decoherence of Landau and Levitov
  Quasiparticles in Quantum Hall Edge Channels}\fi,
 \jr{Physical Review Letters} \textbf{113}, 166403 (2014).


\bibitem{Slobodeniuk2016}% article
 \textsc{A.\,O. Slobodeniuk},  \textsc{E.\,G. Idrisov},  and  \textsc{E.\,V.
  Sukhorukov}\iffalse {Relaxation of an electron wave packet at the quantum
  Hall edge at filling factor $\nu=2$}\fi,
 \jr{Physical Review B} \textbf{93}, 035421 (2016).


\bibitem{Liu1998}% article
 \textsc{R.~Liu},  \textsc{B.~Odom},  \textsc{Y.~Yamamoto},  and
  \textsc{S.~Tarucha}\iffalse {Quantum interference in electron collision}\fi,
 \jr{Nature} \textbf{391}, 263 (1998).


\bibitem{Neder2007}% article
 \textsc{I.~Neder},  \textsc{N.~Ofek},  \textsc{Y.~Chung},
  \textsc{M.~Heiblum},  \textsc{D.~Mahalu},  and  \textsc{V.~Umansky}\iffalse
  {Interference between two indistinguishable electrons from independent
  sources.}\fi,
 \jr{Nature} \textbf{448}, 333 (2007).


\bibitem{Litvin2007}% article
 \textsc{L.~Litvin},  \textsc{H.\,P. Tranitz},  \textsc{W.~Wegscheider},  and
  \textsc{C.~Strunk}\iffalse {Decoherence and single electron charging in an
  electronic Mach-Zehnder interferometer}\fi,
 \jr{Physical Review B} \textbf{75}, 033315 (2007).


\bibitem{Rossello2015}% article
 \textsc{G.~Rossello},  \textsc{F.~Battista},  \textsc{M.~Moskalets},  and
  \textsc{J.~Splettstoesser}\iffalse {Interference and multiparticle effects in
  a Mach-Zehnder interferometer with single-particle sources}\fi,
 \jr{Physical Review B} \textbf{91}, 115438 (2015).


\bibitem{Bertoni2000}% article
 \textsc{A.~Bertoni},  \textsc{P.~Bordone},  \textsc{R.~Brunetti},
  \textsc{C.~Jacoboni},  and  \textsc{S.~Reggiani}\iffalse {Quantum Logic Gates
  based on Coherent Electron Transport in Quantum Wires}\fi,
 \jr{Physical Review Letters} \textbf{84}, 5912 (2000).


\bibitem{Hong1987}% article
 \textsc{C.~Hong},  \textsc{Z.~Ou},  and  \textsc{L.~Mandel}\iffalse
  {Measurement of subpicosecond time intervals between two photons by
  interference}\fi,
 \jr{Physical Review Letters} \textbf{59}, 2044 (1987).


\bibitem{Beugnon2006}% article
 \textsc{J.~Beugnon},  \textsc{M.~Jones},  \textsc{J.~Dingjan},
  \textsc{B.~Darqui{\'{e}}},  \textsc{G.~Messin},  \textsc{A.~Browaeys},  and
  \textsc{P.~Grangier}\iffalse {Quantum interference between two single photons
  emitted by independently trapped atoms.}\fi,
 \jr{Nature} \textbf{440}, 779 (2006).


\bibitem{Flagg2010}% article
 \textsc{E.~Flagg},  \textsc{A.~Muller},  \textsc{S.~Polyakov},
  \textsc{A.~Ling},  \textsc{A.~Migdall},  and  \textsc{G.~Solomon}\iffalse
  {Interference of Single Photons from Two Separate Semiconductor Quantum
  Dots}\fi,
 \jr{Physical Review Letters} \textbf{104}, 4 (2010).


\bibitem{Bocquillon2013a}% article
 \textsc{E.~Bocquillon},  \textsc{V.~Freulon},  \textsc{J.\,M. Berroir},
  \textsc{P.~Degiovanni},  \textsc{B.~Placais},  \textsc{A.~Cavanna},
  \textsc{Y.~Jin},  and  \textsc{G.~Feve}\iffalse {Coherence and
  Indistinguishability of Single Electrons Emitted by Independent Sources}\fi,
 \jr{Science} \textbf{339}, 1054--1057 (2013).


\bibitem{Glauber1963}% article
 \textsc{R.~Glauber}\iffalse {The Quantum Theory of Optical Coherence}\fi,
 \jr{Physical Review} \textbf{130}, 2529--2539 (1963).


\bibitem{Grenier2011NJP}% article
 \textsc{C.~Grenier},  \textsc{R.~Herv{\'{e}}},  \textsc{E.~Bocquillon},
  \textsc{F.\,D. Parmentier},  \textsc{B.~Pla{\c{c}}ais},  \textsc{J.\,M.
  Berroir},  \textsc{G.~F{\`{e}}ve},  and  \textsc{P.~Degiovanni}\iffalse
  {Single-electron quantum tomography in quantum Hall edge channels}\fi,
 \jr{New Journal of Physics} \textbf{13}, 093007 (2011).


\bibitem{Haack2011}% article
 \textsc{G.~Haack},  \textsc{M.~Moskalets},  \textsc{J.~Splettstoesser},  and
  \textsc{M.~B{\"{u}}ttiker}\iffalse {Coherence of single-electron sources from
  Mach-Zehnder interferometry}\fi,
 \jr{Physical Review B} \textbf{84}, 081303 (2011).


\bibitem{Haack2013}% article
 \textsc{G.~Haack},  \textsc{M.~Moskalets},  and
  \textsc{M.~B{\"{u}}ttiker}\iffalse {Glauber coherence of single-electron
  sources}\fi,
 \jr{Physical Review B} \textbf{87}, 201302 (2013).


\bibitem{Moskalets2014}% article
 \textsc{M.~Moskalets}\iffalse {Two-electron state from the Floquet scattering
  matrix perspective}\fi,
 \jr{Physical Review B} \textbf{89}, 045402 (2014).


\bibitem{Thibierge2016}% article
 \textsc{{\'{E}.}.~Thibierge},  \textsc{D.~Ferraro},  \textsc{B.~Roussel},
  \textsc{C.~Cabart},  \textsc{A.~Marguerite},  \textsc{G.~F{\`{e}}ve},  and
  \textsc{P.~Degiovanni}\iffalse {Two-electron coherence and its measurement in
  electron quantum optics}\fi,
 \jr{Physical Review B} \textbf{93}, 081302 (2016).


\bibitem{Ferraro2013}% article
 \textsc{D.~Ferraro},  \textsc{A.~Feller},  \textsc{A.~Ghibaudo},
  \textsc{E.~Thibierge},  \textsc{E.~Bocquillon},  \textsc{G.~F{\`{e}}ve},
  \textsc{C.~Grenier},  and  \textsc{P.~Degiovanni}\iffalse {Wigner function
  approach to single electron coherence in quantum Hall edge channels}\fi,
 \jr{Physical Review B} \textbf{88}, 205303 (2013).


\bibitem{Wigner1932}% article
 \textsc{E.~Wigner}\iffalse {On the Quantum Correction For Thermodynamic
  Equilibrium}\fi,
 \jr{Physical Review} \textbf{40}, 749 (1932).


\bibitem{Smithey1993}% article
 \textsc{D.~Smithey},  \textsc{M.~Beck},  \textsc{M.~Raymer},  and
  \textsc{A.~Faridani}\iffalse {Measurement of the Wigner distribution and the
  density matrix of a light mode using optical homodyne tomography: Application
  to squeezed states and the vacuum}\fi,
 \jr{Physical Review Letters} \textbf{70}, 1244--1247 (1993).


\bibitem{Bertet2002}% article
 \textsc{P.~Bertet},  \textsc{A.~Auffeves},  \textsc{P.~Maioli},
  \textsc{S.~Osnaghi},  \textsc{T.~Meunier},  \textsc{M.~Brune},
  \textsc{J.\,M. Raimond},  and  \textsc{S.~Haroche}\iffalse {Direct
  Measurement of the Wigner Function of a One-Photon Fock State in a
  Cavity}\fi,
 \jr{Physical Review Letters} \textbf{89}, 200402 (2002).


\bibitem{Mahe2010}% article
 \textsc{A.~Mah{\'{e}}},  \textsc{F.~Parmentier},  \textsc{E.~Bocquillon},
  \textsc{J.\,M. Berroir},  \textsc{D.~Glattli},  \textsc{T.~Kontos},
  \textsc{B.~Pla{\c{c}}ais},  \textsc{G.~F{\`{e}}ve},  \textsc{A.~Cavanna},
  and  \textsc{Y.~Jin}\iffalse {Current correlations of an on-demand
  single-electron emitter}\fi,
 \jr{Physical Review B} \textbf{82}, 201309 (2010).


\bibitem{Albert2010}% article
 \textsc{M.~Albert},  \textsc{C.~Flindt},  and
  \textsc{M.~B{\"{u}}ttiker}\iffalse {Accuracy of the quantum capacitor as a
  single-electron source}\fi,
 \jr{Physical Review B} \textbf{82}, 41407 (2010).


\bibitem{Jonckheere2012a}% article
 \textsc{T.~Jonckheere},  \textsc{T.~Stoll},  \textsc{J.~Rech},  and
  \textsc{T.~Martin}\iffalse {Real-time simulation of finite-frequency noise
  from a single-electron emitter}\fi,
 \jr{Physical Review B} \textbf{85}, 045321 (2012).


\bibitem{Parmentier2012}% article
 \textsc{F.\,D. Parmentier},  \textsc{E.~Bocquillon},  \textsc{J.\,M. Berroir},
   \textsc{D.\,C. Glattli},  \textsc{B.~Pla{\c{c}}ais},
  \textsc{G.~F{\`{e}}ve},  \textsc{M.~Albert},  \textsc{C.~Flindt},  and
  \textsc{M.~B{\"{u}}ttiker}\iffalse {Current noise spectrum of a
  single-particle emitter: Theory and experiment}\fi,
 \jr{Physical Review B} \textbf{85}, 165438 (2012).


\bibitem{Bocquillon2012}% article
 \textsc{E.~Bocquillon},  \textsc{F.\,D. Parmentier},  \textsc{C.~Grenier},
  \textsc{J.\,M. Berroir},  \textsc{P.~Degiovanni},  \textsc{D.\,C. Glattli},
  \textsc{B.~Pla{\c{c}}ais},  \textsc{A.~Cavanna},  \textsc{Y.~Jin},  and
  \textsc{G.~F{\`{e}}ve}\iffalse {Electron Quantum Optics: Partitioning
  Electrons One by One}\fi,
 \jr{Physical Review Letters} \textbf{108}, 196803 (2012).


\bibitem{Olkhovskaya2008}% article
 \textsc{S.~Ol'khovskaya},  \textsc{J.~Splettstoesser},  \textsc{M.~Moskalets},
   and  \textsc{M.~B{\"{u}}ttiker}\iffalse {Shot Noise of a Mesoscopic
  Two-Particle Collider}\fi,
 \jr{Physical Review Letters} \textbf{101}, 166802 (2008).


\bibitem{Lee1997}% article
 \textsc{H.~Lee} and  \textsc{S.~Yang}\iffalse {Spin-charge separation in
  Quantum Hall Liquids}\fi,
 \jr{Physical Review B} \textbf{56}, R15529--R15532 (1997).


\bibitem{Pham2000}% article
 \textsc{K.\,V. Pham},  \textsc{M.~Gabay},  and  \textsc{P.~Lederer}\iffalse
  {Fractional excitations in the Luttinger liquid}\fi,
 \jr{Physical Review B} \textbf{61}(24), 16397--16422 (2000).


\bibitem{Berg2009}% article
 \textsc{E.~Berg},  \textsc{Y.~Oreg},  \textsc{E.\,A. Kim},  and
  \textsc{F.~von Oppen}\iffalse {Fractional Charges on an Integer Quantum Hall
  Edge}\fi,
 \jr{Physical Review Letters} \textbf{102}, 236402 (2009).


\bibitem{Kamata2010}% article
 \textsc{H.~Kamata},  \textsc{T.~Ota},  \textsc{K.~Muraki},  and
  \textsc{T.~Fujisawa}\iffalse {Voltage-controlled group velocity of edge
  magnetoplasmon in the quantum Hall regime}\fi,
 \jr{Physical Review B} \textbf{81}, 085329 (2010).


\bibitem{Kumada2011}% article
 \textsc{N.~Kumada},  \textsc{H.~Kamata},  and  \textsc{T.~Fujisawa}\iffalse
  {Edge magnetoplasmon transport in gated and ungated quantum Hall systems}\fi,
 \jr{Physical Review B} \textbf{84}, 045314 (2011).


\bibitem{Kamata2014}% article
 \textsc{H.~Kamata},  \textsc{N.~Kumada},  \textsc{M.~Hashisaka},
  \textsc{K.~Muraki},  and  \textsc{T.~Fujisawa}\iffalse {Fractionalized wave
  packets from an artificial Tomonaga?Luttinger liquid}\fi,
 \jr{Nature Nanotechnology} \textbf{9}, 177--181 (2014).


\bibitem{Kovrizhin2011}% article
 \textsc{D.~Kovrizhin} and  \textsc{J.~Chalker}\iffalse {Equilibration of
  integer quantum Hall edge states}\fi,
 \jr{Physical Review B} \textbf{84}, 085105 (2011).


\bibitem{Kovrizhin2012}% article
 \textsc{D.~Kovrizhin} and  \textsc{J.~Chalker}\iffalse {Relaxation in driven
  integer quantum Hall edge states}\fi,
 \jr{Physical Review Letters} \textbf{109}, 106403 (2012).


\bibitem{Levkivskyi2009}% article
 \textsc{I.~Levkivskyi} and  \textsc{E.~Sukhorukov}\iffalse {Noise-Induced
  Phase Transition in the Electronic Mach-Zehnder Interferometer}\fi,
 \jr{Physical Review Letters} \textbf{103}, 036801 (2009).


\bibitem{Levkivskyi2012}% article
 \textsc{I.~Levkivskyi} and  \textsc{E.~Sukhorukov}\iffalse {Energy relaxation
  at quantum Hall edge}\fi,
 \jr{Physical Review B} \textbf{85}, 075309 (2012).


\bibitem{Freulon2015}% article
 \textsc{V.~Freulon},  \textsc{A.~Marguerite},  \textsc{J.\,M. Berroir},
  \textsc{B.~Pla{\c{c}}ais},  \textsc{A.~Cavanna},  \textsc{Y.~Jin},  and
  \textsc{G.~F{\`{e}}ve}\iffalse {Hong-Ou-Mandel experiment for temporal
  investigation of single-electron fractionalization}\fi,
 \jr{Nature Communications} \textbf{6}, 6854 (2015).


\bibitem{Bocquillon2013b}% article
 \textsc{E.~Bocquillon},  \textsc{V.~Freulon},  \textsc{J.\,M. Berroir},
  \textsc{P.~Degiovanni},  \textsc{B.~Pla{\c{c}}ais},  \textsc{A.~Cavanna},
  \textsc{Y.~Jin},  and  \textsc{G.~F{\`{e}}ve}\iffalse {Separation of neutral
  and charge modes in one-dimensional chiral edge channels}\fi,
 \jr{Nature Communications} \textbf{4}, 1839 (2013).


\bibitem{Jonckheere2012}% article
 \textsc{T.~Jonckheere},  \textsc{J.~Rech},  \textsc{C.~Wahl},  and
  \textsc{T.~Martin}\iffalse {Electron and hole Hong-Ou-Mandel
  interferometry}\fi,
 \jr{Physical Review B} \textbf{86}, 125425 (2012).


\bibitem{Marguerite2016}% article
 \textsc{A.~Marguerite},  \textsc{C.~Cabart},  \textsc{C.~Wahl},
  \textsc{B.~Roussel},  \textsc{V.~Freulon},  \textsc{D.~Ferraro},
  \textsc{C.~Grenier},  \textsc{J.\,M. Berroir},  \textsc{B.~Pla{\c{c}}ais},
  \textsc{T.~Jonckheere},  \textsc{J.~Rech},  \textsc{T.~Martin},
  \textsc{P.~Degiovanni},  \textsc{A.~Cavanna},  \textsc{Y.~Jin},  and
  \textsc{G.~F{\`{e}}ve}\iffalse {Decoherence and relaxation of a single
  electron in a one dimensional conductor}\fi,
 \jr{to appear in PRB, arXiv 1609.03494} (2016).


\bibitem{LeSueur2010}% article
 \textsc{H.~{Le Sueur}},  \textsc{C.~Altimiras},  \textsc{U.~Gennser},
  \textsc{A.~Cavanna},  \textsc{D.~Mailly},  and  \textsc{F.~Pierre}\iffalse
  {Energy relaxation in the integer quantum Hall regime.}\fi,
 \jr{Physical Review Letters} \textbf{105}, 056803 (2010).


\bibitem{Altimiras2010}% article
 \textsc{C.~Altimiras},  \textsc{H.~{Le Sueur}},  \textsc{U.~Gennser},
  \textsc{A.~Cavanna},  \textsc{D.~Mailly},  and  \textsc{F.~Pierre}\iffalse
  {Tuning Energy Relaxation along Quantum Hall Channels}\fi,
 \jr{Physical Review Letters} \textbf{105}, 226804 (2010).


\bibitem{Inoue2013}% article
 \textsc{H.~Inoue},  \textsc{A.~Grivnin},  \textsc{N.~Ofek},
  \textsc{I.~Neder},  \textsc{M.~Heiblum},  \textsc{V.~Umansky},  and
  \textsc{D.~Mahalu}\iffalse {Charge Fractionalization in the Integer Quantum
  Hall Effect}\fi,
 \jr{Physical Review Letters} \textbf{112}, 166801 (2014).


\othercit
\bibitem{Wiener1949}% book
 \textsc{N.~Wiener},
{Extrapolation, Interpolation, and Smoothing of Stationary Time Series} (Wiley
  and Sons, 1949).


\bibitem{Dubois2013}% article
 \textsc{J.~Dubois},  \textsc{T.~Jullien},  \textsc{C.~Grenier},
  \textsc{P.~Degiovanni},  \textsc{P.~Roulleau},  and  \textsc{D.\,C.
  Glattli}\iffalse {Integer and fractional charge Lorentzian voltage pulses
  analyzed in the framework of photon-assisted shot noise}\fi,
 \jr{Physical Review B} \textbf{88}, 085301 (2013).


\bibitem{Jullien2014}% article
 \textsc{T.~Jullien},  \textsc{P.~Roulleau},  \textsc{B.~Roche},
  \textsc{A.~Cavanna},  \textsc{Y.~Jin},  and  \textsc{D.~Glattli}\iffalse
  {Quantum tomography of an electron}\fi,
 \jr{Nature} \textbf{514}, 603--607 (2014).


\bibitem{Hofer2013}% article
 \textsc{P.~Hofer} and  \textsc{M.~B{\"{u}}ttiker}\iffalse {Emission of
  time-bin entangled particles into helical edge states}\fi,
 \jr{Physical Review B} \textbf{88}, 241308 (2013).


\bibitem{Inhofer2013}% article
 \textsc{A.~Inhofer} and  \textsc{D.~Bercioux}\iffalse {Proposal for an
  on-demand source of polarized electrons into the edges of a topological
  insulator}\fi,
 \jr{Physical Review B} \textbf{88}, 235412 (2013).


\bibitem{Ferraro2014a}% article
 \textsc{D.~Ferraro},  \textsc{C.~Wahl},  \textsc{J.~Rech},
  \textsc{T.~Jonckheere},  and  \textsc{T.~Martin}\iffalse {Electronic
  Hong-Ou-Mandel interferometry in two-dimensional topological insulators}\fi,
 \jr{Physical Review B} \textbf{89}, 075407 (2014).


\bibitem{Ferraro2014b}% article
 \textsc{D.~Ferraro},  \textsc{J.~Rech},  \textsc{T.~Jonckheere},  and
  \textsc{T.~Martin}\iffalse {Nonlocal interference and Hong-Ou-Mandel
  collisions of single Bogoliubov quasiparticles}\fi,
 \jr{Physical Review B} \textbf{91}, 075406 (2015).


\bibitem{Calzona2015}% article
 \textsc{A.~Calzona},  \textsc{M.~Carrega},  \textsc{G.~Dolcetto},  and
  \textsc{M.~Sassetti}\iffalse {Time-resolved pure spin fractionalization and
  spin-charge separation in helical Luttinger liquid based devices}\fi,
 \jr{Physical Review B} \textbf{92}, 195414 (2015).


\bibitem{Dolcetto2016}% article
 \textsc{G.~Dolcetto} and  \textsc{T.~Schmidt}\iffalse {Emission of entangled
  Kramers pairs from a helical mesoscopic capacitor}\fi,
 \jr{Physical Review B} \textbf{94}, 075444 (2016).


\bibitem{Roussel2016}% article
 \textsc{B.~Roussel},  \textsc{C.~Cabart},  \textsc{G.~F{\`{e}}ve},
  \textsc{E.~Thibierge},  and  \textsc{P.~Degiovanni}\iffalse {Electron quantum
  optics as quantum signal processing}\fi,
 \jr{To appear in the present volume}.


\end{thebibliography}

\end{document}